\documentclass[twocolumn,trackchanges]{aastex631}
\usepackage{gensymb}
\usepackage{amsmath}

\shorttitle{Evidence of Gas Depletion}
\shortauthors{Wen et al.}

\begin{document}

\title{Evidence of Gas Depletion in Quasars with Moderate Radio Emission}

\author[0009-0005-9436-1142]{Yuhan Wen}
\affiliation{Department of Astronomy, School of Physics, Peking University, 
Beijing 100871, People's Republic of China}
\affiliation{Kavli Institute for Astronomy and Astrophysics, Peking University, 
Beijing 100871, People's Republic of China}

\author[0000-0003-4956-5742]{Ran Wang}
\affiliation{Department of Astronomy, School of Physics, Peking University, 
Beijing 100871, People's Republic of China}
\affiliation{Kavli Institute for Astronomy and Astrophysics, Peking University, 
Beijing 100871, People's Republic of China}

\author[0000-0001-6947-5846]{Luis C. Ho}
\affiliation{Department of Astronomy, School of Physics, Peking University, 
Beijing 100871, People's Republic of China}
\affiliation{Kavli Institute for Astronomy and Astrophysics, Peking University, 
Beijing 100871, People's Republic of China}

\author[0000-0002-4569-9009]{Jinyi Shangguan}
\affiliation{Department of Astronomy, School of Physics, Peking University, 
Beijing 100871, People's Republic of China}
\affiliation{Kavli Institute for Astronomy and Astrophysics, Peking University, 
Beijing 100871, People's Republic of China}

\author[0000-0001-7568-6412]{Ezequiel Treister}
\affiliation{Instituto de Alta Investigaci{\'{o}}n, Universidad de Tarapac{\'{a}}, Casilla 7D, Arica, Chile}

\author[0000-0003-4007-5771]{Guodong Li}
\affiliation{Department of Astronomy, School of Physics, Peking University, 
Beijing 100871, People's Republic of China}
\affiliation{Kavli Institute for Astronomy and Astrophysics, Peking University, 
Beijing 100871, People's Republic of China}

\author[0000-0002-8686-8737]{Franz E. Bauer}
\affiliation{Instituto de Alta Investigaci{\'{o}}n, Universidad de Tarapac{\'{a}}, Casilla 7D, Arica, Chile}

\begin{abstract}

The energy released by active galactic nuclei (AGNs) is considered to have a profound impact on the cold gas properties of their host galaxies, potentially heating or removing the gas and further suppressing star formation. To understand the feedback from AGN radio activity, we investigate its impacts on the cold gas reservoirs in AGNs with different radio activity levels. We construct a quasar sample with a mean $z\sim1.5$ and a mean $L_{\rm bol}\sim10^{45.8}\ \rm erg\ s^{-1}$, all with Herschel detections to enable estimates of the total gas mass through the galactic dust continuum emission. The sample is then cross-matched with radio catalogs and divided into radio loud (RL) quasars, radio-detected radio quiet (RQ) quasars and radio-undetected quasars based on their radio loudness. Through spectral energy distribution (SED) fitting, we find the radio-detected RQ quasars exhibit evidence of gas deficiency with host galaxies possessing $\sim 0.3$ dex lower dust and gas masses compared to the other two groups, despite being matched in $M_{\rm BH}$, $L_{\rm bol}$, $M_{*}$ and SFR. Furthermore, evidence from optical spectra shows that both the fraction and velocity of outflows are higher in the radio-detected RQ group, suggesting a connection between the ionized gas outflows and the moderate radio activity. These results suggest that the AGN feedback could be more efficient in AGNs with weak/moderate radio emission than in those without radio detection or those with strong radio emission. Further high-resolution observations are needed to understand the interaction between the interstellar medium and the weak/moderate AGN radio activity.

\end{abstract}


\section{Introduction} \label{sec:intro}

Active Galactic Nuclei (AGNs), one of the most energetic phenomena in the universe, are powered by the accretion of matter onto supermassive black holes (SMBHs). While the gravitational influence of SMBHs is confined to the innermost regions of galaxies, the vast energy generated through accretion processes largely exceeds the binding energy of the galaxy bulge \citep{Fabian2012}. The released energy, if efficiently coupled with the surrounding gas, can have a profound impact on the host galaxy by heating and/or expelling the gas of the interstellar medium (ISM). Such processes, known as AGN feedback, can operate in various forms, such as radiation pressure, winds, and radio jets \citep[see][for a review]{Fabian2012}, playing a crucial role in regulating star formation and leading to the AGN-galaxy coevolution scenario \citep{SilkRees1998, Kormendy2013}.

AGN feedback can be divided into two main modes, the radiative mode (or quasar/wind mode) and the kinetic mode (or radio mode) \citep{Heckman2014}. In the radiative mode, the radiation from AGN accretion disk can drive powerful winds, heating or removing surrounding gas, potentially suppressing star formation by preventing gas from cooling and collapsing \citep[e.g.,][]{Cicone2014, Harrison2014}. The kinetic mode, on the other hand, is typically associated with radio emission. In most cases, the radio emission is characterized by low power, poor collimation and compact structures, which could inject significant energy into the ISM and thereby regulate star formation \citep[e.g.,][]{IllustrisTNG}. A minority of AGNs host powerful and highly-collimated jets, which are able to drive energetic outflows in host galaxies \citep[e.g.,][]{Nesvadba2008, Brusa2016, Vayner2017}. Other studies also suggest these jets to be capable of heating the circumgalactic medium (CGM) and preventing gas from cooling \citep[e.g.,][]{McNamara2012}. Due to their distinct impact, feedback from high-power jets is sometimes classified separately as the jet mode.

Although AGN feedback is proposed to play an important role in SMBH and galaxy co-evolution \citep[e.g.,][]{IllustrisTNG, SIMBA}, observational evidence regarding its impact on the gas content and distribution in AGN host galaxies remains controversial. %In the local Universe, 
Previous studies have found abundant evidence of high velocity outflows on galactic scales in AGN host galaxies, in the form of both ionized gas and molecular gas \citep[e.g.,][]{Feruglio2010, Cicone2014, Harrison2014, Brusa2016}. These outflows are often interpreted as direct evidence of AGN feedback. Several studies on samples with $z\simeq1-5$ also suggested that AGN host galaxies are more gas-depleted than non-active galaxies (e.g., based on X-ray selected AGNs, \citealt{Bertola2024}; mid-infrared (MIR) selected quasars, \citealt{Bischetti2021}). However, other studies on samples with $z\sim0.01$ revealed that AGN host galaxies exhibit similar global molecular gas properties compared with non-active galaxies with similar stellar masses, such as molecular gas mass and star formation efficiency (e.g., based on X-ray selected Seyfert galaxies, \citealt{Rosario2018}; MIR selected Seyfert galaxies, \citealt{Salvestrini2022}). Moreover, by comparing the kinetic pressure to the gravitational pressure in local AGN host galaxies, detailed kinematic studies have revealed that the ISM is in a state of hydrostatic equilibrium, similar to that in normal star-forming galaxies, implying minimal AGN feedback effects on the molecular gas content in their host galaxies \citep{Fei2023, Fei2024}.
%At higher redshifts, 
%Nevertheless, other studies found marginal differences in molecular gas properties between AGNs and non-active galaxies, implying minimal AGN feedback effects on their host galaxies \citep[e.g.,][]{Circosta2021}. 

Despite extensive research on AGN feedback, the role of radio activity has not been thoroughly discussed. Previous studies mainly focus on the radio-loud AGNs, as powerful radio jets are suggested to be linked with strong outflows \citep[e.g.,][]{Holt2008,Nesvadba2008,Kim2013}. \citet{Best2005} investigated the host galaxy properties of radio-loud AGNs and found they share similar properties to those of ordinary galaxies, with a tendency for radio-loud AGNs to be found in larger galaxies and in richer environments. Some studies discussed the outflow properties in AGNs with detectable radio emission, finding the width of the [O {\sc iii}] $\lambda$5007 line is the broadest in AGNs with moderate radio luminosity \citep[$L_{\text{1.4 GHz}}\sim10^{24} \rm{\ W\ Hz^{-1}}$,][]{Mullaney2013}. However, the impact of radio emission on the total gas reservoirs of host galaxies in AGNs with different radio activity levels remains poorly understood. 

In this work, we aim to explore the effects of AGN feedback on host galaxies across a broad range of radio activity levels, from radio-undetected to radio-detected AGNs and from radio-quiet to radio-loud AGNs. Specifically, we investigate how different levels of radio activity influence the cold gas content in host galaxies. By doing so, we hope to provide new insights into the complex interplay between AGN activity and galaxy evolution.

To quantify the gas content in AGN host galaxies, we utilize sub-millimeter (sub-mm) observations ($\sim250-870\mathrm{\mu m}$ in the observed frame) to measure the galactic dust emission, which serves as a reliable proxy of cold gas \citep{Saintonge2022}. Probing the Rayleigh-Jeans tail at these long wavelengths ensures that the signal is dominated by cold dust and is free from significant AGN contamination \citep[e.g.,][]{Kirkpatrick2012}. Furthermore, we account for potential contributions from AGN radio emission by incorporating a radio model into our SED fitting process (Section \ref{sec:analysis}). By constructing an AGN sample with Herschel/SPIRE detections and cross-matching this AGN sample with radio catalogs, we investigate different impacts of AGN feedback on the gas content in AGNs with different radio activity levels. The paper is organized as follows. Our sample selection and data collection are described in Section \ref{sec:samp}. Data analysis, including SED fitting and spectral fitting is described in Section \ref{sec:analysis}. We present the distributions of the physical properties obtained by the SED fitting and the spectral fitting in Section \ref{sec:result}. We discuss our results with respect to AGN feedback and coevolution in Section \ref{sec:discuss} and conclude in Section \ref{sec:sum}. We adopt a $\Lambda$CDM cosmology with $\Omega_m=0.287, \Omega_\Lambda=0.713$, and $H_0=69.32 \rm\ km\ s^{-1}\ Mpc^{-1}$ \citep{WMAP9}.

\section{Sample Selection and Data Collection} \label{sec:samp}

We construct a quasar sample selected from the Sloan Digital Sky Survey \citep[SDSS,][]{SDSS} Stripe 82 region. SDSS Stripe 82 is a region covering about 270 deg$^2$ along the celestial equator in the Southern Galactic Cap. Stripe 82 is also well known for its abundant multi-wavelength data from X-ray to radio bands. Most importantly, there are existing sub-mm surveys \citep[e.g., HerS,][]{HerS} and radio surveys \citep[e.g., VLA 1.4GHz survey,][]{VLAS82} in this region, providing valuable information about the dust content and radio properties of quasars.

Our sample selection follows \citet{DongWu16}, who studied a far infrared (FIR) bright quasar sample consisting of 207 quasars selected by cross-matching the SDSS quasar catalogs [seventh data release (DR7Q) and tenth data release (DR10Q)] with the HerS survey catalog. Based on this sample of 207 quasars, we expand the sample to 436 quasars by cross-matching the new SDSS quasar catalog sixteenth data release \citep[DQ16Q,][]{SDSSDR16Q} with the HerS catalog. 

\begin{figure}
    \centering
    \includegraphics[width=0.8\linewidth]{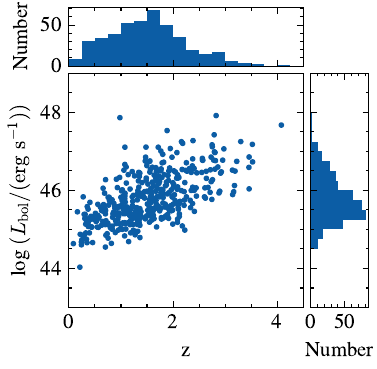}
    \caption{The distribution of the redshift and $L_{\rm bol}$ of our sample \citep[from][]{SDSSDR16Q_pro}. The upper panel and the right panel show the distributions of the redshift and $L_{\rm bol}$ respectively.}
    \label{fig:Lbol-z}
\end{figure}

Following \citet{DongWu16}, we adopt a matching radius of 5 arcsec and find 258 new targets in total. To ensure the reliability of the quasar redshift, we compare the redshifts provided by SDSS DR16Q with those of SDSS DR18 \citep{SDSSDR18}, and discard 12 sources with redshift difference greater than 0.1 and 1 source without spectra data. After further examination of the spectra through spectral fitting (see Section \ref{subsec:spec} for details), we discard 5 sources with unreliable redshifts. Another 3 sources are discarded due to blending and 2 sources are discarded due to large radio variability, which are discussed in Sections \ref{subsec:opt} and \ref{subsec:radio}, respectively, in detail. We further exclude 6 sources based on ALMA observations that show no detection at the quasar position, but show clear signals from surrounding sources, indicating that the Herschel flux densities are therefore dominated by these surrounding emitters (Section \ref{subsubsec:fir}).\citet{SDSSDR16Q_pro} provide AGN bolometric luminosity ($L_{\rm bol}$) and monochromatic luminosities (at 1350 \text{\AA}, 3000 \text{\AA} and 5100 \text{\AA}) for the DR16Q sample. As shown in Figure \ref{fig:Lbol-z}, our sample spans redshifts from 0.08 to 4.1 and $L_{\rm bol}$ from $10^{44}$ to $10^{48} \ \mathrm{erg~s^{-1}}$, with a mean redshift of 1.5 and a mean $L_{\rm bol}$ of $10^{45.8} \rm{\ erg\ s^{-1}}$.

\subsection{Optical Photometric Data and Spectroscopic Data} \label{subsec:opt}

The SDSS imaging survey includes five optical bands ($ugriz$) and provides various magnitude measurements. Due to the large redshift range from 0.08 to 4.1, our sample consists of both unresolved and extended sources. The PSF magnitudes are recommended for unresolved point sources, but fail to account for the outer regions of extended sources. On the contrary, the cModel magnitudes work better for extended sources, but may introduce additional uncertainties for point sources. Therefore, using only the PSF magnitude or the cModel magnitude is insufficient for the total sample.

\begin{figure*}[tbh!]
    \centering
    \includegraphics[width=0.7\linewidth]{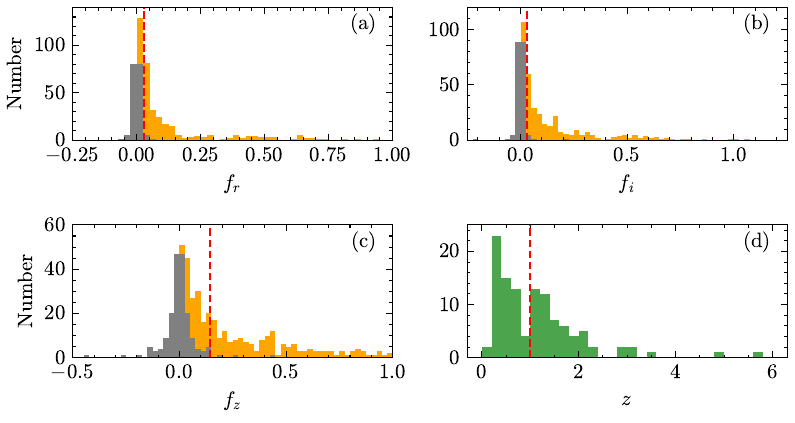}
    \caption{(a)-(c) Distributions of the relative flux difference $f_b\equiv (F_{b,\rm cModel}-F_{b,\rm PSF})/{F_{b,\rm cModel}}$ in $r$, $i$ and $z$ bands for the total sample (orange) and the symmetric distribution (gray) obtained by mirroring the negative-$f_b$ part to the positive side.} The red dashed line indicates the 95th percentile of the mirrored positive distribution, above which the sources are considered extended. (d) The redshift distribution of the subsample selected with $f_r>0.027,\ f_i>0.031,$ and $f_z>0.144$. The red dashed line marks the gap at $z=1$, and we require the extended sample to lie below this redshift.
    \label{fig:fb_dist}
\end{figure*}

To quantify the difference between the PSF flux and the cModel flux, we define
\begin{equation}
    f_b\equiv \frac{F_{b,\rm cModel}-F_{b,\rm PSF}}{F_{b,\rm cModel}},
\end{equation}
where $F_{\rm cModel}$ and $F_{\rm PSF}$ refer to the PSF flux and the cModel flux respectively, and $b$ refers to any band in $ugriz$ bands. For point sources, $f_b$ is expected to be around 0, while for extended sources, $f_b$ is expected to be positive. Thus, this parameter can be used to distinguish extended sources from point sources. Figure \ref{fig:fb_dist} shows the distributions of $f_r$, $f_i$ and $f_z$ for our sample. We focus on the $r$, $i$ and $z$ bands for this analysis due to their smaller PSF widths compared to the $u$ and $g$ bands\footnote{https://www.sdss4.org/dr15/imaging/}.

From Figure \ref{fig:fb_dist}, the extended sources appear as a long positive tail in the distribution of $f_b$ in each band. We assume that the negative side of the $f_b$ distribution reflects the intrinsic measurement uncertainties of point sources. Therefore, to model this uncertainty distribution, we extract the negative part of each distribution and mirror it to the positive side, shown as the gray histogram. We determine the thresholds of $f_r$, $f_i$ and $f_z$ for extended sources as the 95th percentile of the positive side of the gray histogram in each band, shown as the red dashed line, i.e., $f_r>0.027,\ f_i>0.031,$ and $f_z>0.144$. 

The redshift distribution of the objects satisfying the above criteria is shown in the bottom right panel of Figure \ref{fig:fb_dist}, where a clear gap appears at $z\sim1$. Quasars at $z>1$ can hardly be resolved by SDSS and the differences between the PSF flux and the cModel flux may result from large flux uncertainties and/or source blending. Therefore, we consider sources with $f_r>0.027,\ f_i>0.031,\ f_z>0.144$ and $z<1$ as extended, while the rest are point sources. We further examine the image cutout of objects at $z>1$ in Figure \ref{fig:fb_dist}d using higher-resolution data from Hyper Suprime-Cam \citep[HSC,][]{HSCPDR3} and Dark Energy Spectroscopic Instrument \citep[DESI,][]{DESIlegacy}. This leads to the removal of three clearly blended systems (SDSS J010307.97+011407.0, SDSS J011227.87-003151.6, and SDSS J013834.18-000509.3) from the subsequent analysis. The remaining $z>1$ objects are retained in the point source category. Based on the above classification, we adopt the PSF magnitudes for point sources and the cModel magnitudes for extended sources.

In addition to the photometric data, we also utilize SDSS optical spectroscopic data to measure the BH mass and ionized gas outflow properties (Section \ref{subsec:spec}). SDSS optical spectroscopy is obtained with SDSS-I / II spectrographs for DR7 data (wavelength coverage $\sim 3800 - 9200$ \text{\AA}) and upgraded BOSS spectrographs \citep{BOSS} for SDSS-III, SDSS-IV and SDSS-V data (wavelength coverage $\sim 3650 - 10400$ \text{\AA}). The spectral resolving power R is about 2000.

\subsection{UV Data} \label{subsec:UV}

The Galaxy Evolution Explorer \citep[GALEX,][]{GALEX} is a satellite that observed galaxies in the near-UV (NUV) and far-UV (FUV) wavelengths from July 2003 to February 2012. Following the methods in \citet{GALEXcounter}, we match quasars with GALEX tiles that simultaneously have positive effective exposure time in both NUV and FUV, i.e., $t_{\rm NUV}>1\ \rm s$ and $t_{\rm FUV}>1\ \rm s$. We also require that the positions of our sample are within the central $0.5 \degree$ radius of the field of view (FOV) of any GALEX tile, i.e., FOV offset $\leq 0.5 \degree$. We search for the GALEX counterparts to our sample within a matching radius of $2.6''$ as suggested by \citet{GALEX_crossradius}. For targets with more than one measurement, we choose the one with the longest FUV exposure time.

The full width half maximum (FWHM) of the PSF of the GALEX image is $\sim 4.2''$ in FUV and $\sim 5.3''$ in NUV. As the magnitudes given in GALEX catalogs are measured in an elliptical Kron aperture, the flux measurements could be contaminated if there are nearby stars or galaxies. We manually inspect the image cutouts from GALEX and SDSS and discard GALEX measurements that are contaminated due to nearby sources.

\subsection{Infrared Data} \label{subsec:IR}

\subsubsection{Near-IR}
The UKIRT Infrared Deep Sky Survey \citep[UKIDSS,][]{UKIDSS} is a near-infrared (NIR) sky survey, carrying Y ($1.03\ \mu \rm m$), J ($1.25\ \mu \rm m$), H ($1.63\ \mu \rm m$) and K ($2.20\ \mu \rm m$) bands, with a K-band depth to 18.3 mag. We cross-match our sample with the UKIDSS DR11PLUS Large Area Survey (LAS) catalog using a matching radius of $2''$ \citep{Roseboom2013}. For the point sources defined in Section \ref{subsec:opt}, we adopt the aperture-corrected magnitude AperMag3 ($2.0''$ aperture diameter) as suggested in \citet{UKIDSSEDR}. For the extended sources, we adopt the PetroMag \citep[Petrosian aperture,][]{Petro} that accounts for extended emission from host galaxies.

\begin{figure*}[htb!]
    \centering
    \includegraphics[width=0.6\linewidth]{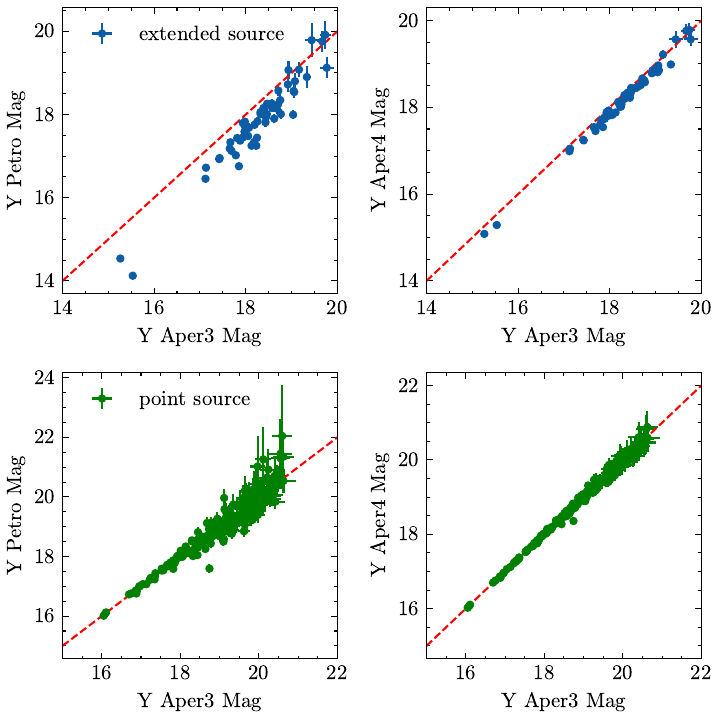}
    \caption{The comparisons between different magnitudes of UKIDSS. The blue points in top panels refer to the magnitudes of extended sources. The green points in bottom panels refer to the magnitudes of point sources. The red dashed line in each plot is the one-to-one line.}
    \label{fig:UKIDSS_magcompare}
\end{figure*}

We further examine the classification into point sources and extended sources described in Section \ref{subsec:opt} by comparing different magnitude measurements, as shown in Figure \ref{fig:UKIDSS_magcompare}. The point sources show great consistency among AperMag3, AperMag4 ($2.8''$ aperture diameter) and PetroMag, while the extended sources have smaller PetroMag and AperMag4 compared to AperMag3 as expected. This comparison proves the reliability of our classification.

\subsubsection{Mid-IR}

The Wide-field Infrared Survey Explorer \citep[WISE,][]{WISE} maps the whole sky in 4 infrared bands W1, W2, W3 and W4 centered at 3.4, 4.6, 12 and 22 $\mu\rm m$, with an angular resolution of $6.1'',\ 6.4'',\ 6.5''$ and $12.0''$ respectively. We cross-match our sample with the AllWISE source catalog \citep{AllWISE} using a matching radius of $3''$ \citep{Roseboom2013} and adopt the profile-fitting magnitudes. The AllWISE catalog also provides a $95\%$ confidence upper limit when the flux measurement has SNR $<2$. We adopt the upper limit for sources with SNR $<2$ to constrain the SED at Mid-IR wavelength.

\subsubsection{Far-IR/Submillimeter} \label{subsubsec:fir}

The Herschel Stripe 82 survey \citep[HerS,][]{HerS} is a sub-mm survey covering $79 \rm\ deg^2$ along the SDSS Stripe 82 region, achieving an average depth of 13.0, 12.9 and 14.8 $\rm mJy\ beam^{-1}$ at 250, 350, 500 $\mu \rm m$. The FWHM of the PSF is $18.15'',\ 25.15''$ and $36.3''$ respectively and the confusion limit is $\sim 7\rm\ mJy$ in three bands. For the mean redshift of our sample, the observed wavelengths correspond to the rest-frame FIR, tracing the cold dust emission. The source photometry of HerS is performed using a modified version of the De-blended SPIRE Photometry (DESPHOT) algorithm, which deals with the source deblending issue in a way appropriate to SPIRE maps. The HerS catalog includes sources with SNR $>3$ at 250 $\mu \rm m$ and provides the noise measurements considering both the instrumental noise and the confusion noise. As suggested in \citet{DongWu16}, we adopt a matching radius of $5''$. To obtain a larger sample, we do not require the target to be detected in all three Herschel bands. In the final sample, all targets are nevertheless detected in at least two Herschel bands (i.e., 250 and 350 $\mu\rm{m}$).

We use the Atacama Large Millimeter/submillimeter Array (ALMA) archive data to better constrain the dust emission at longer wavelengths. We searched the ALMA archive and found a subset of our sample was observed in Band 7 as part of Project 2016.2.00060.S \citep[PI: E. Hatziminaoglou,][]{Hatziminaoglou2018} and Project 2022.1.00029.S \citep[PI: E. Hatziminaoglou,][]{Hatziminaoglou2025}. All the targets in our sample observed in the former project were also observed in the latter one, with higher resolution and better sensitivity. Therefore, we use the data from Project 2022.1.00029.S, which proposed ALMA 12-m array band 7 continuum observations at a resolution of $\sim0.8''$ of a statistical, FIR-bright sample of 153 SDSS quasars at 343 GHz (870 $\mu\rm m$). The data were observed in October 2022 and released in January 2024. 

We downloaded the pipeline-reduced image data which were produced with Briggs weighting and robust parameter equal to 0.5 and by combining all 4 spectral windows (SPWs). We use the Common Astronomy Software Applications \citep[CASA,][]{CASA} task {\tt imfit} to fit the source. There are a total of 135 detected sources in our sample. Under the spatial resolution, 54 sources are resolved, 22 sources are marginally resolved (resolved along one axis) and the remaining 59 sources are unresolved. For a consistent measurement across our sample, we adopt the integrated flux density given by {\tt imfit} task for following analysis. We note that for another 6 sources observed by ALMA, no emission was detected at the quasar positions, while clear signals are detected from surrounding sources. This indicates that the observed Herschel flux is likely dominated by these surrounding sources rather than the quasars themselves. Therefore, we discard these 6 sources in the following analysis. However, we caution that a similar situation may exist for other sources in our sample without ALMA observations. Given that such cases constitute only a small fraction ($\sim4\%$) of the ALMA-observed subsample, we consider the potential contamination to be minor and expect it to have no significant impact on our results.

\subsection{Radio Data} \label{subsec:radio}

\subsubsection{VLA Stripe 82 Survey and FIRST Survey}

The Very Large Array (VLA) Stripe 82 survey \citep{VLAS82} is a high-resolution 1.4 GHz survey along the SDSS Stripe 82 region, achieving an angular resolution of $1.8''$ and a median rms noise of 52 $\mu\rm Jy\ beam^{-1}$ over 92 $\rm deg^2$. We adopt a matching radius of $1''$ as recommended in \citet{VLAS82} and find 43 detected sources in our sample. A few sources of the HerS catalog ($\sim8\%$ in our sample) lie outside the VLA Stripe 82 region. Therefore, we further cross-match our sample with the Faint Images of the Radio Sky at Twenty-cm \citep[FIRST,][]{FIRST} survey using a matching radius of $1.5''$ \citep{Ivezi2002} and find four new sources that are not detected by VLA Stripe 82 survey. FIRST is also a survey conducted with VLA at 1.4 GHz with an angular resolution of $5''$ and a typical rms of 0.15 $\rm mJy$. We use the integrated flux density $F_{\rm int}$ instead of the peak flux density $F_{\rm peak}$ to account for possible extended radio structures in our sample. We note that while the FIRST survey has a different depth from the VLA Stripe 82 survey, the sources supplemented in this way constitute a very small fraction of our total sample, and their inclusion does not change the main results presented in this work.

The catalogs of both VLA Stripe 82 survey and FIRST survey only include sources brighter than $5\sigma$. To enlarge the sample size with radio detection, we further examine the image cutout of VLA Stripe 82 survey and FIRST survey, and measure the peak flux density of the image. We find 14 new sources brighter than $3\sigma$ and within $1''$ of their optical counterparts. Because none of these sources are resolved by VLA, the peak flux density can be regarded as the integrated flux density for following analysis. For sources brighter than $5\sigma$ in VLA Stripe 82 catalog, we also examine the image cutout and find most are unresolved with the angular resolution of $1.8''$, corresponding to $5\sim15$ kpc in physical scale.

\subsubsection{VLASS}

The Very Large Array Sky Survey \citep[VLASS,][]{VLASS} is a 3 GHz survey conducted using VLA with an angular resolution of $\sim 2.5''$ and a 1$\sigma$ sensitivity goal of 120 $\mu\rm Jy$ for a single epoch. The survey will map 80 percent of the sky in three epochs over seven years and is expected to catalog approximately 10 million radio sources. VLASS provides two types of products: the Quick Look (QL) image product and the Single Epoch (SE) image product. The QL image products are produced as soon as possible ($\sim2-4$ weeks) after observations, while the SE image products take more time due to algorithmic difficulties and compute needs. Due to the limited sky coverage of the current SE catalog, we use the QL Epoch 2 catalog for cross-matching and adopt a matching radius of $1.5''$ \citep{Zhong2025}, finding 12 detected sources in our sample.

We also examine the image cutout of VLASS and measure the peak flux density of the image. We find 6 new sources brighter than $3\sigma$ and within $1.5''$ of their optical counterparts. Additionally, we find 1 target (SDSS J005905.51+000651.6) lying outside the region of Epoch 2, but observed in Epoch 3. We downloaded the image cutout of this target and use the CASA task {\tt imfit} to fit the source, yielding an integrated flux density of $1.699\pm0.021 \rm\ Jy$ and a peak flux density of $1.537\pm0.011 \rm\ Jy\ beam^{-1}$. We correct for the systematic under-measurement of flux density values in the QL images \citep{VLASS_under} before performing SED fitting.

\subsubsection{Other Radio Data} \label{subsubsec:otherradio}

In total, we find 60 sources detected in radio surveys, out of which 18 objects are detected at both 1.4 GHz and 3 GHz, 41 objects are only detected at 1.4 GHz, and 1 object is only detected at 3 GHz. We estimate the radio loudness for objects with radio detection using the traditional definition, $R=L_{\rm 5 GHz}/L_{4400\mathring{\rm A}}$ \citep{Kellerman1989}. We estimate the flux density at rest-frame 5 GHz assuming a power-law radio spectrum $f_{\nu}\propto \nu^{-\alpha}$ and a spectral index of $\alpha=0.7$ for quasars with only one radio-band observation \citep{deGasperin2018}. For quasars with both 1.4 GHz and 3 GHz radio data, we extrapolate to estimate the flux density at rest-frame 5 GHz. $L_{4400\mathring{\rm A}}$ is calculated using the monochromatic luminosities from \citet{SDSSDR16Q_pro}. For each quasar, we determine the spectral index from the closest available bands assuming a power-law continuum, and then interpolate to $4400\mathring{\rm A}$.

Seven quasars in our sample have radio loudness larger than 1000. For these quasars, we further search for other archive radio data at NASA/IPAC Extragalactic Database (NED). Among the seven radio extremely loud quasars, SDSS J012528.84-000555.9 and SDSS J021605.65-011803.5 exhibit flat spectra at low frequency but steep spectra at high frequency. For a better decomposition of the radio emission and the dust emission at sub-mm bands, we use the 229 GHz and 86 GHz IRAM 30-m telescope observations for J0125 \citep{J0125_radio}, and use 43.34 GHz and 22.46 GHz VLA observations for J0216 \citep{J0216_radio}. Another notable quasar, SDSS J010838.76+013500.2, shows large variability in a 850 $\mu \rm m$ monitoring observation \citep{Jenness2010}. Due to large radio variability, the difference in sub-mm flux densities for J0108 from the HerS catalog and from the Herschel SPIRE \citep{SPIRE} catalog is also significant (the flux density is $\sim 50\text{ mJy}$ larger in the SPIRE catalog). A similar situation occurs for SDSS J005905.51+000651.6, which shows a large flux density difference between the two catalogs. Therefore, we discard J0108 and J0059 in the following analysis, as we are unable to obtain reliable dust emission due to the non-simultaneous observation time. The remaining four quasars either show a consistent spectral index from low to high frequencies or are not observed at other radio bands.

\section{Data Analysis} \label{sec:analysis}

We now describe our methodology for the modeling of the broad-band SED and the spectra of our sample.

\subsection{SED Fitting} \label{subsec:SED}

We employ the Code Investigating GALaxy Emission (CIGALE) algorithm \citep{CIGALE_Boquien,CIGALE_Yang20,CIGALE_Yang22} to perform the SED fitting to the photometric data described above. CIGALE is an efficient SED fitting code written in Python language, which can model the spectra of AGNs from X-ray to radio wavelengths and estimate the physical properties (e.g., star formation rate (SFR), stellar mass ($M_*$) and dust mass ($M_{\rm dust}$)). CIGALE carefully considers the AGN contribution by incorporating the SKIRTOR AGN torus model \citep{SKIRTOR}, which handles clumpy dust distributions and polar dust extinction. The dust emission is also taken into account through the energy balance between the UV/optical and mid-/far- IR wavelengths.

The models used in this paper are as follows. The galaxy component is modeled using a delayed Star Formation History (SFH) model {\tt sfhdelayed} with a functional form ${\rm SFR}(t)\propto t/\tau^2\times\exp(-t/\tau)$. The model also allows for an exponential burst representing the latest episode of star formation. The stellar population is modeled using the single stellar population templates of \citet{BC03}. The dust attenuation model {\tt dustatt\_modified\_starburst} is based on the \citet{Calzetti} starburst attenuation curve, extended with the \citet{Leitherer} curve between the Lyman break and 150 nm. We adopt the dust emission model {\tt dl2014} to model the emission from dust heated by stars. The model generally follows \citet{Draine2007} and is updated by \citet{Draine2014}. The AGN emission is incorporated using the {\tt skirtor2016} model. We also include the radio model for sources with radio data, which accounts for the radio emission from both the star formation and the AGN. The parameter space of each model is presented in Table \ref{tab:SED_para}. The models and parameter settings are mainly based on \citet{Magnelli2023, reducedchi2}, where the former study on star-forming galaxies and the latter study focuses on local AGNs from SDSS. We have adopted a wider range of {\tt fracAGN} to better account for the high-luminosity quasars in our sample. Parameters not listed in Table \ref{tab:SED_para} are set to their CIGALE default values. Specifically, in the {\tt skirtor2016} model, we adopt an opening angle of $40\degree$ and an outer-to-inner radius ratio of 20 for the torus component, and adopt a dust temperature of 100 K and a dust emissivity index of 1.6 for the polar dust component.

\begin{deluxetable*}{ccc}[htb!]
\centering
\tablecaption{Models and parameter space used for the CIGALE SED fitting of our sample. \label{tab:SED_para}}
\tablehead{
\colhead{Parameter} & \colhead{Description} & \colhead{Value} \\
}
\startdata
\multicolumn{3}{l}{SFH: {\tt sfhdelayed}}\\ 
\hline
$\tau$\_main  & e-folding time of the main population & 500, 2000, 5000, 8000 Myr\\
age\_main     & Age of the main population & 500, 2000, 5000, 8000 Myr\\
f\_burst      & Mass fraction of the late burst population & 0, 0.05, 0.1, 0.2\\ 
\hline
\multicolumn{3}{l}{Stellar emission: {\tt bc03} \citep{BC03}}\\ 
\hline
IMF           & Initial mass function & Chabrier\\
Z             & Metallicity & 0.02\\ 
\hline
\multicolumn{3}{l}{Dust attenuation: {\tt dustatt\_modified\_starburst} \citep{Calzetti}}\\ 
\hline
E\_BV\_lines  & Color excess of the nebular lines & 0.01, 0.1, 0.3, 0.5, 0.7, 0.9, 1.2\\
E\_BV\_factor & Reduction factor to apply on E\_BV\_lines & 0.25, 0.75\\ 
\hline
\multicolumn{3}{l}{Dust emission: {\tt dl2014} \citep{Draine2014}}\\ 
\hline
umin          & Minimum radiation field & 0.1, 1, 10, 30, 50\\
$\gamma$      & Fraction illuminated from Umin to Umax & 0.2, 0.1, 0.02, 0.001\\ 
\hline
\multicolumn{3}{l}{AGN model: {\tt skirtor2016} \citep{SKIRTOR}}\\ 
\hline
t             & Average edge-on optical depth at 9.7 micron & 3, 7, 11\\
i             & Inclination & 0, 20, 40 (type 1); 50, 60, 80 (type 2) degrees\\
fracAGN       & AGN fraction & 0.05, 0.1, 0.2, 0.3, 0.4, 0.5, 0.6, 0.7, 0.8, 0.9, 0.95\\
EBV           & E(B-V) for the extinction in the polar direction in magnitudes & 0, 0.03, 0.2\\ 
\hline
\multicolumn{3}{l}{Radio emission: {\tt radio}}\\ 
\hline
R\_agn        & The radio-loudness parameter for AGN & 0.1, 1, 5, 10, 50, 100, 200, 500, 1000,\\
{           } & { } & 2000, 5000, 10000, 20000, 50000 \\
$\alpha$\_agn & The slope of the power-law AGN radio emission & 0, 0.3, 0.5, 0.7, 1.2 (only for J0125)                                        
\enddata
\end{deluxetable*}

The radio model of CIGALE considers the AGN radio emission using a single power-law over the wavelength range of $0.1-1000 \rm\ mm$, without considering different spectral indices at different wavelengths. For the two extremely radio loud objects J0125 and J0216, the radio spectra are getting steeper at higher frequency. Therefore, we use only the measurement at higher frequencies for a better decomposition of the radio emission and the dust emission (see Section \ref{subsubsec:otherradio} for details). For J0125, we find the upper limit of the radio spectral index for other sources (i.e., $\rm max(\alpha_{\rm agn})=0.7$) cannot well describe the radio spectrum of this source. Therefore, we calculate the spectral index using the measurements at 229 GHz and 86 GHz and fix it at 1.2 when performing SED fitting to J0125.

From the SED fitting, we obtained the SFR and $M_*$ of the quasar host galaxies, based on a Chabrier initial mass function \citep[IMF,][]{Chabrier2003}. We note that different choices of IMF would introduce a systematic offset in SFR and $M_*$ estimates. For example, adopting a Salpeter IMF \citep{Salpeter1955} would increase these values by $\sim0.2$ dex \citep{Madau2014}. Furthermore, due to the degeneracy between the stellar and AGN components, the estimates of $M_*$ may be subject to systematic uncertainties of up to 0.17 dex, depending on the AGN fraction \citep{Ciesla2015}. On the contrary, the SFRs may be less sensitive to the degeneracy as they are better constrained with the FIR data, i.e., the Herschel observations. However, we caution that the SFRs of some targets in our sample may be overestimated due to the low resolution of Herschel, which may include contamination from surrounding sources (e.g., companion galaxies, foreground galaxies). The contamination will be further discussed in Section \ref{subsec:SEDresult}.

Assuming an optically thin case, we measure the dust masses using
\begin{equation}
    M_{\rm dust}=\frac{S_{1200\mu \rm m}D_L^2}{(1+z)\kappa_{0,1200\mu \rm m}B_\nu(T_{\rm dust})},
\end{equation}
where $S_{1200\mu \rm m}$ is the observed flux density at rest-frame 1200 $\mu$m fitted by CIGALE, $D_L$ is the luminosity distance, $\kappa_{0,1200\mu\rm m}$ is the mass absorption coefficient at 1200 $\mu$m, $B_\nu$ is the Planck function, and $T_{\rm dust}$ is the dust temperature. We adopt $\kappa_{0,1200\mu \rm m}=0.4\rm{\ cm^2\ g^{-1}}$ and $T_{\rm dust}=40\rm\ K$ \citep{Beelen2006,Alton2004}. The choice of the rest-frame 1200 $\mu$m ensures that we probe the Rayleigh–Jeans tail of the dust emission, where the emission is optically thin, and that AGNs are not expected to contribute significantly to the emission at this wavelength. We do not adopt the dust mass directly derived by CIGALE, as it assumes a typical gas-to-dust ratio of Milky Way to estimate dust mass, which may not be applicable to our sample. Another widely-used method to estimate the dust mass is to fit the dust continuum with a modified blackbody \citep[MBB, e.g.,][]{Casey2012}. We discuss possible systematic offsets between dust masses derived from different methods in Appendix \ref{app:dust}. When the dust emissivity index is fixed at $\beta=1.6$, the dust masses obtained from two methods are consistent. However, when $\beta=2$ is adopted, the MBB-derived dust masses are 0.25 dex lower than those from CIGALE. Despite these systematic offsets, the relative differences in $M_{\rm dust}$ remain significant across groups with different radio levels. Further details are discussed in Appendix \ref{app:dust}.

The gas masses are estimated by
\begin{equation}
    M_{\rm gas}=M_{\rm HI}+M_{\rm H_2}=\delta_{\rm GDR}\ M_{\rm dust},
\end{equation}
where $\delta_{\rm GDR}$ is the gas-to-dust ratio, which is a function of the gas-phase metallicity \citep{Draine2007, Leroy2011}. Following \citet{Shangguan18}, we adopt the $\delta_{\rm GDR}-Z$ relation recalibrated by \citet{GDRtoZ}
\begin{equation}
\begin{aligned}
    \log \delta_{\rm GDR}=\ &(10.54\pm1.0)-(0.99\pm0.12)\\
    &\times[12+\log(\rm O/H)]
\end{aligned}
\end{equation}
with a scatter of 0.15 dex, and the $M_*-Z$ relation given by \citet{MstartoZ}
\begin{equation}
\begin{aligned}
    12+\log({\rm O/H})=\ &23.9049-5.62784\log M_*\\
    &+0.645142\times(\log M_*)^2\\
    &-0.0235065\times(\log M_*)^3
\end{aligned}
\end{equation}
with a residual scatter of 0.09 dex.

\subsection{Spectral Fitting} \label{subsec:spec}

\begin{figure*}[htb!]
    \centering
    \includegraphics[width=0.8\linewidth]{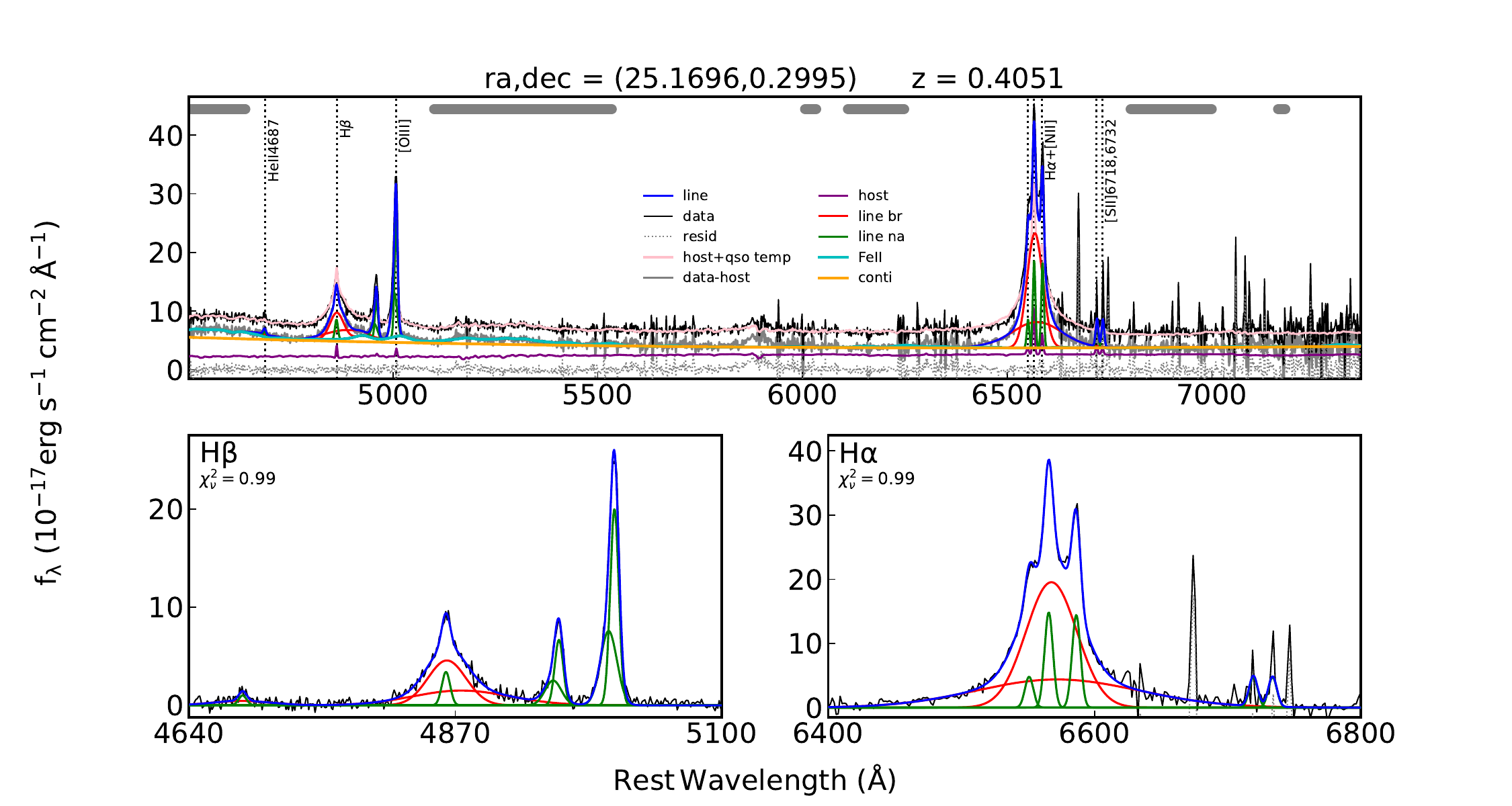 }
    \caption{The spectral fitting results of SDSS J014040.71+001758.1 with the spectral fitting code PyQSOFit. 
    \label{fig:spec_example}}
\end{figure*}

Apart from modeling the photometric data using CIGALE, we adopt the spectral fitting code PyQSOFit \citep{PyQSOFit,PyQSOFit_Ren,PyQSOFit_Shen19} to decompose the SDSS spectra of our sample and measure spectral properties of emission lines. PyQSOFit first fits the continuum in given continuum windows that consists of the power-law emission from the accretion disk, the optical and UV Fe {\sc ii} templates, and a host galaxy component. Then the continuum is subtracted to fit the emission lines. We use two Gaussians to fit the broad components of emission lines and use one Gaussian to fit the narrow components. For [O {\sc iii}] 4959 \text{\AA} and 5007 \text{\AA} emission lines, we use one Gaussian to fit any asymmetric wing component and one Gaussian to fit the core component, with the wing component tracing the outflow of the ionized gas. We present an example of the spectral fitting in Figure \ref{fig:spec_example}.

The Bayesian Information Criterion (BIC) can be used to determine whether broad components are needed to fit the emission lines. Using this criterion, we identify 6 type 2 quasars and 11 quasars with broad components in H$\alpha$ emission lines but only narrow components in H$\beta$. The remaining 419 sources are type 1 quasars. The type information is used to constrain the inclination angle of the AGN model during SED fitting. The inclination angle values for type 1 quasars include 0, 20 and 40 degrees, while the values for type 2 quasars include 50, 60, and 80 degrees.

Based on the single epoch virial BH mass estimator \citep[e.g.,][]{VP06}, we estimate BH masses from the AGN disk continuum and broad emission line measurements for type 1 quasars. The virial mass estimate can be expressed as:
\begin{equation}
\begin{aligned}
    \log \left(\frac{M_{\rm BH,vir}}{M_\odot}\right)=&\ a+b\times\log\left(\frac{\lambda L_\lambda}{10^{44}\ \rm erg\ s^{-1}}\right)\\
    &+c\times\log\left(\frac{\rm FWHM}{\rm km\ s^{-1}}\right),
\end{aligned}
\end{equation}
where $\lambda L_\lambda$ refers to the AGN continuum luminosity at wavelength $\lambda$, FWHM is the measured full width at half-maximum of the total broad line profile, coefficients a, b, and c are calibrated using local AGNs with reverberation mapping BH mass measurements. The coefficients used in this work for different broad emission lines are listed in Table \ref{tab:coeff}.

\begin{deluxetable}{cccccc}[]
\tablecaption{Coefficients of single epoch $M_{\rm BH}$ estimation for different broad lines. \label{tab:coeff}}
\tablehead{
\colhead{Line} & \colhead{$\lambda_{\rm conti}$} & \colhead{a} & \colhead{b} & \colhead{c} & \colhead{Reference} \\
}
\startdata 
H$\alpha$ & 5100 & 0.789 & 0.53 & 2.06 & GH05, HK15 \\
H$\beta$  & 5100 & 0.910 & 0.53 & 2    & HK15 \\
Mg {\sc ii}     & 3000 & 0.740 & 0.62 & 2    & S11 \\
C {\sc iv}      & 1450 & 0.660 & 0.53 & 2    & VP06
\enddata
\tablecomments{$\lambda_{\rm conti}$ refers to the wavelength of the continuum luminosity used for each emission line.}
\tablerefs{GH05 \citep{Greene_Ha}, HK15 \citep{Ho2015}, S11 \citep{Shen11_MgII}, VP06 \citep{VP06}.}
\end{deluxetable}

We only estimate the BH masses for quasars with the broad emission line SNR $>2$ (defined as peak signal/rms). We set a loose lower limit (FWHM $\sim$ 800 km/s) to the width of the broad component to account for a possible contribution from the outer part of the broad line region (BLR). However, we exclude the emission lines with a total FWHM smaller than 1200 km/s when estimating the central BH mass using the single epoch spectrum. We note that it sets a lower limit to the BH mass, which is $\sim10^{7.5}~M_\odot$ given the mean $L_{\rm bol}$ of our sample. For quasars with more than one emission line available, we adopt the BH mass according to the priority order of H$\alpha$, H$\beta$, Mg {\sc ii}, and C {\sc iv}.

\section{Results} \label{sec:result}

\subsection{Host Galaxy Properties} \label{subsec:SEDresult}

\begin{figure}[htb!]
    \centering
    \includegraphics[width=\linewidth]{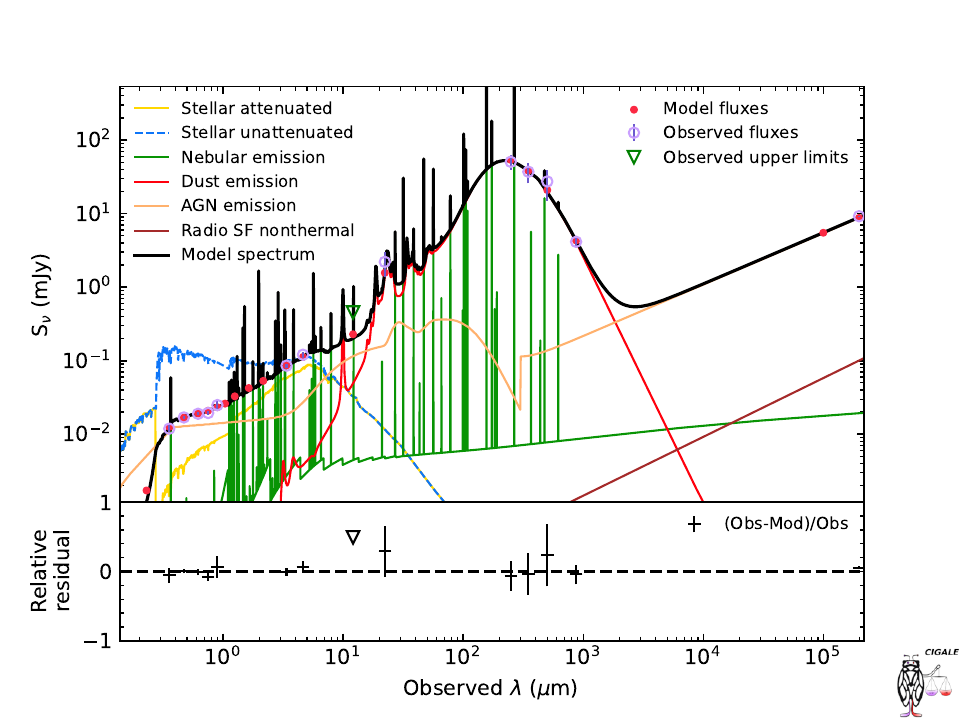}
    \includegraphics[width=\linewidth]{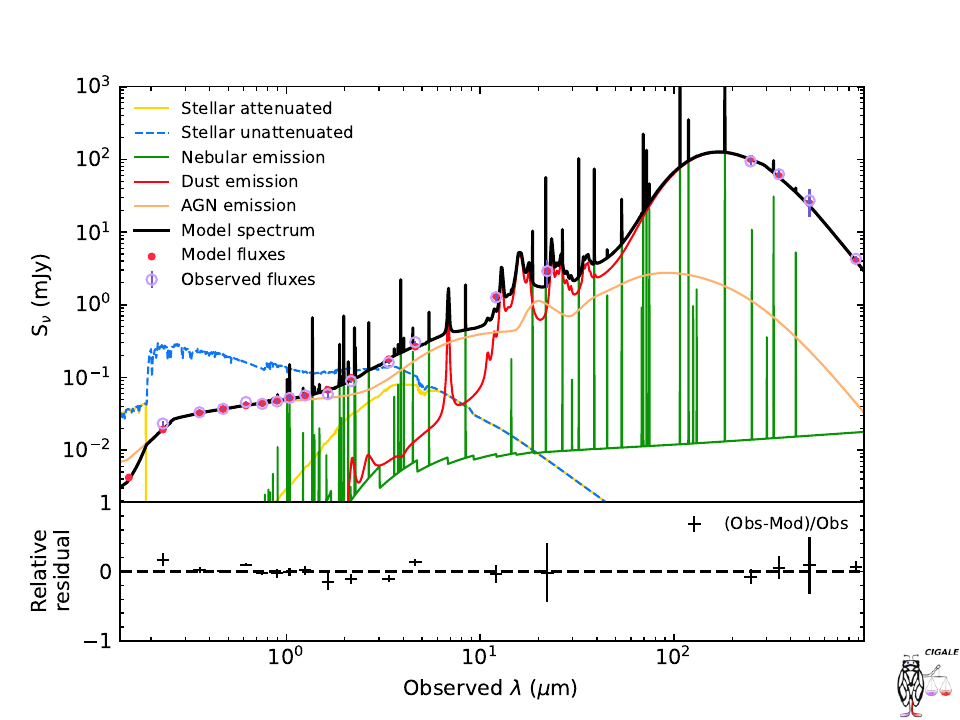}
    \caption{The SED fitting results of two examples, SDSS J021916.41-012100.0 (top) and SDSS J005743.78-001158.9 (bottom), constructed using the {\tt CIGALE v2022.0} code. 
    \label{fig:SED_example}}
\end{figure}

To ensure the reliability of the SED fitting results, we exclude 3 quasars with reduced $\chi^2$ greater than 10, i.e. $\chi^2_\nu>10$. We present two examples of the SED fitting results in Figure \ref{fig:SED_example}, one with radio detection and one without. Table \ref{tab:five} shows the first five entries of the estimated physical properties of our sample.

\begin{deluxetable*}{ccccccccccc}[htb!]
\centering
\tablecaption{First five entries of basic information and physical properties of our quasar sample. \label{tab:five}}
\tablehead{
\colhead{SDSS ID} & \colhead{redshift} & \colhead{log SFR} & \colhead{log $M_*$} & \colhead{log $M_{\rm dust}$} & \colhead{log $M_{\rm gas}$} & \colhead{log $M_{\rm BH}$} & \colhead{log $L_{\rm bol}$} & \colhead{R} \\
\colhead{} & \colhead{} & \colhead{($M_\odot\ \rm yr^{-1}$)} & \colhead{($M_\odot$)} & \colhead{($M_\odot$)} & \colhead{($M_{\odot}$)} & \colhead{($M_\odot$)} & \colhead{($\rm erg\ s^{-1}$)} & \colhead{}
}
\startdata
005624.62-000438.5 & 2.24 & 2.90$\pm$0.19 & 11.06$\pm$0.28 & 8.47$\pm$0.19 & 10.32$\pm$1.46 & 8.77$\pm$0.22 & 46.516$\pm$0.003 & $<$4.87 \\
005642.28+000104.7 & 1.08 & 2.66$\pm$0.04 & 10.81$\pm$0.20 & 8.97$\pm$0.18 & 10.84$\pm$1.46 & $\cdots$ & 45.861$\pm$0.005 & 4.39$\pm$0.83 \\
005713.03+005725.5 & 1.77 & 2.55$\pm$0.18 & 10.68$\pm$0.31 & 8.77$\pm$0.20 & 10.64$\pm$1.46 & 8.84$\pm$0.20 & 46.503$\pm$0.003 & $<$2.78 \\
005743.78-001158.9 & 1.07 & 2.67$\pm$0.12 & 10.75$\pm$0.20 & 8.74$\pm$0.18 & 10.61$\pm$1.46 & 8.13$\pm$0.31 & 45.787$\pm$0.009 & $<$4.68 \\
005828.93+001346.8 & 1.24 & 2.42$\pm$0.21 & 11.40$\pm$0.21 & 8.33$\pm$0.21 & 10.20$\pm$1.47 & 8.56$\pm$0.20 & 45.874$\pm$0.007 & $<$9.10 \\
\enddata
\tablecomments{The columns are as follows: SDSS Name, redshift, star formation rate, stellar mass, dust mass, gas mass, BH mass, bolometric luminosity and radio loudness. For radio-undetected quasars, a $3\sigma$ upper limit for the radio loudness is presented, which is calculated using the local rms noise from the VLA Stripe 82 survey or alternatively from FIRST when unavailable, and adopting a radio spectral index of $\alpha=0.7$. This table is available in its entirety in machine-readable form.
}
\end{deluxetable*}

With the measured SFR, we estimate the radio emission contributed from star formation using the relation between SFR and $L_{\rm 1.4GHz}$ \citep{Hopkins2003,Suresh2024}. Our analysis of the 62 radio-detected quasars reveals that 90\% exhibit radio luminosities exceeding those expected from star formation alone. Moreover, in 63\% of these quasars, the star formation contribution accounts for less than 50\% of the total radio luminosity, suggesting a dominant role of AGN in radio activity.

We divide the sample into three groups based on the radio loudness: the radio loud (RL) group, the VLA-detected radio quiet (RQ) group and the radio-undetected group. The RL group consists of quasars with $R>10$ and the VLA-detected RQ group consists of quasars with $R\leq10$. The VLA-detected RQ group, though with low radio loudness, is still detected in radio bands, indicating moderate radio emission in these quasars.

\begin{figure*}[htb!]
    \centering
    \includegraphics[width=0.7\linewidth]{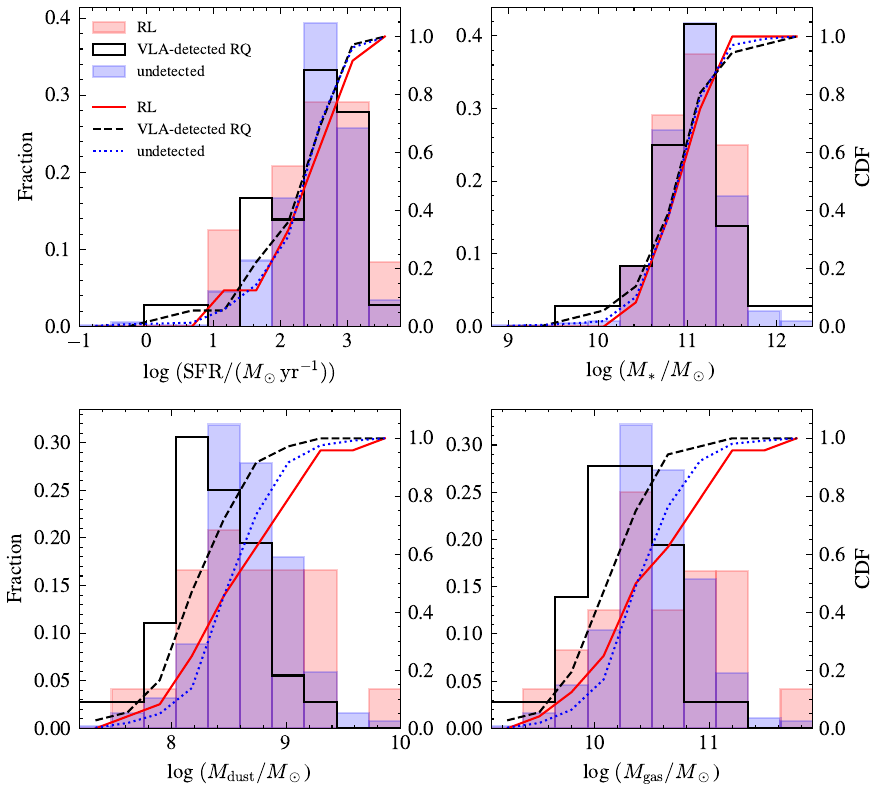}
    \caption{The distributions (histogram, left y-axis) and cumulative distribution functions (line, right y-axis) of SFRs, $M_*$, $M\rm _{dust}$ and $M\rm _{gas}$ of the RL group (shown in red), the VLA-detected RQ group (shown in black) and the radio-undetected group (shown in blue).}
    \label{fig:hist_init}
\end{figure*}

The distributions of the host galaxy properties of three groups are shown in Figure \ref{fig:hist_init}. From the distributions, the SFR and $M_*$ distributions of the three groups are consistent, while the VLA-detected RQ group has $\sim 0.3$ dex lower $M\rm _{dust}$ and $M\rm _{gas}$. We perform statistical tests including the KS test and the t test, and present the results in Table \ref{tab:stats}. The statistical tests are performed as pairwise comparisons among the radio-undetected, the VLA-detected RQ and the RL groups. Table \ref{tab:stats} also includes a comparison with control samples, the construction of which is described below. From the p values in the table, the differences in $M\rm _{dust}$ and $M\rm _{gas}$ between the VLA-detected RQ group and the radio-undetected group are statistically significant ($p\lesssim10^{-4}$). The differences between the VLA-detected RQ group and the RL group are less significant ($0.01<p<0.05$), which may result from the small sample size of radio-detected quasars. 

Again, we caution that due to the low spatial resolution of Herschel observations, the measured Herschel flux densities for some targets in our sample may be contaminated by surrounding sources, potentially leading to over-estimated SFRs. Similarly, the $M_{\rm dust}$ could also be overestimated for targets without ALMA detection. However, when ALMA observations are available, $M_{\rm dust}$ can be robustly estimated, as the ALMA observations not only provide flux densities at the Rayleigh-Jeans tail but also offer sufficient resolution to isolate emission from the central quasar. Nevertheless, such contamination from surrounding sources is unlikely to bias the statistics when making comparisons among different groups, provided that the contamination rate is similar across groups. Based on ALMA observations, we identify possible contamination in 24 out of 114 radio-undetected quasars, 4 out of 16 VLA-detected RQ quasars, and 3 out of 5 RL quasars. Although the ALMA-detected radio-loud sample is too small for definitive conclusions, the contamination fraction is consistent between the radio-undetected and VLA-detected RQ groups. This supports the assumption of a comparable contamination rate across different groups, indicating that such contamination should not significantly bias the relative comparisons among the three groups.

\begin{figure*}[htb!]
    \centering
    \includegraphics[width=0.7\linewidth]{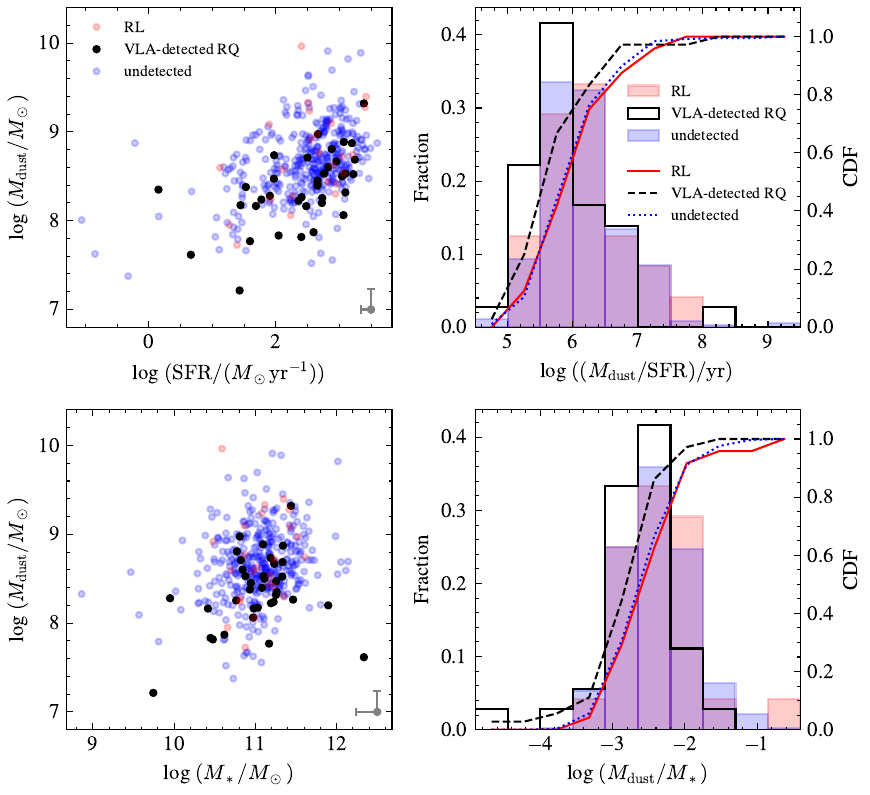}
    \caption{\textit{Left}: $M_{\rm dust}$ vs. SFR and $M_{\rm dust}$ vs. $M_{*}$ of the RL group (red dot), the VLA-detected RQ group (black dot) and the radio-undetected group (blue dot). The typical uncertainty is presented in the lower right corner. \textit{Right}: The distributions (histogram, left y-axis) and cumulative distribution functions (line, right y-axis) of $M_{\rm dust}/{\rm SFR}$ and $M_{\rm dust}/M_{*}$ with color coding the same as Figure \ref{fig:hist_init}.}
    \label{fig:tdepfdust}
\end{figure*}

We further investigate the relations between $M_{\rm dust}$ and SFR, as well as $M_{\rm dust}$ and $M_*$, along with the distributions of $M_{\rm dust}/{\rm SFR}$ and the dust fraction $f_{\rm dust}\ (=M_{\rm dust}/M_*)$. The ratio $M_{\rm dust}/{\rm SFR}$ is expected to trace the gas depletion time, under the assumption of a constant gas-to-dust ratio. We mainly focus on the dust mass instead of the gas mass because large uncertainties are introduced when estimating the gas mass using the gas-to-dust ratio. As shown in Figure \ref{fig:tdepfdust}, the distributions of quasars on both the $M_{\rm dust}-\rm SFR$ plane and the $M_{\rm dust}-M_*$ plane show positive correlations, though with large scatters. Given that dust masses generally scale proportionally to the overall gas masses, these correlations are broadly expected from the integrated Kennicutt–Schmidt law \citep{Kennicutt2021} and the molecular gas Main Sequence \citep{Tacconi2020}. From the histograms on the right panel and the statistical results in Table \ref{tab:stats}, the distributions of $M_{\rm dust}/{\rm SFR}$ and $f_{\rm dust}$ show significant differences between the VLA-detected RQ group and the radio-undetected group ($p\lesssim0.05$), while they show minimal differences between the VLA-detected RQ group and the RL group, possibly due to limited sample size.

\begin{figure*}[htb!]
    \centering
    \includegraphics[width=0.7\linewidth]{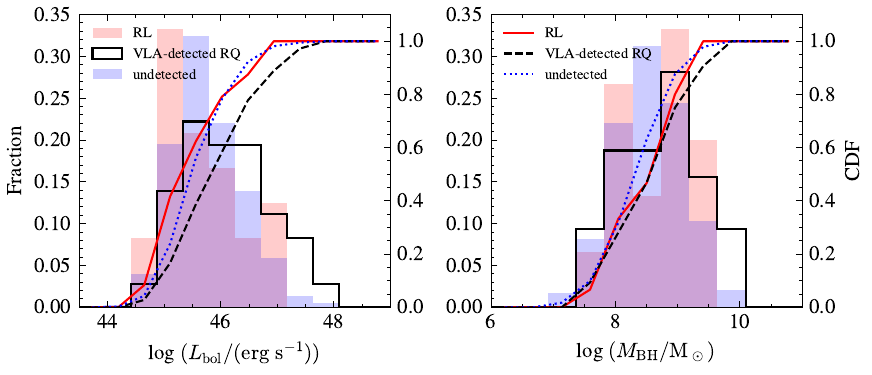}
    \caption{The distributions (histogram, left y-axis) and cumulative distribution functions (line, right y-axis) of $L_{\rm bol}$ and $M_{\rm BH}$ with color coding the same as Figure \ref{fig:hist_init}.}
    \label{fig:LbolMBH}
\end{figure*}

To obtain more reliable results, we restrict the $\chi^2_\nu$ of the SED fitting to $\chi^2_\nu< 5$ \citep{reducedchi2}, and require the source to have a reliable BH mass measurement (with broad emission line SNR $>2$). We further check the distributions of $L_{\rm bol}$ and $M_{\rm BH}$ across the three groups, as shown in Figure \ref{fig:LbolMBH} and Table \ref{tab:stats}. Both t-test and KS-test indicate that the VLA-detected RQ group has slightly higher $L_{\rm bol}$ than the radio-undetected group, with $p<0.05$. Compared to the RL group, the VLA-detected RQ group also shows higher $L_{\rm bol}$, with t-test $p<0.05$ and KS-test $p=0.14$.

To avoid the influence of $L_{\rm bol}$ and $M_{\rm BH}$, we perform two acceptance-rejection resamplings in the radio-undetected group, requiring that the distributions of $M_{\rm BH}$ and $L_{\rm bol}$ from the two resamplings are consistent with those of the RL group and the VLA-detected RQ group, respectively. The sample size of each control sample is set to 100. The results of the statistical tests are listed in Table \ref{tab:stats}. The VLA-detected RQ group still shows a significant difference in $M_{\rm dust}$ and $M_{\rm gas}$ from the radio-undetected group ($p<0.01$), while the RL group shows consistent results with the radio-undetected group. In addition, we note that the difference in $M_{\rm dust}/\rm SFR$ between the VLA-detected RQ group and the radio-undetected group becomes statistically insignificant after resampling, indicating that the previously observed significance was likely introduced by the difference in $L_{\rm bol}$ between the groups. The difference in $f_{\rm dust}$ remains significant ($p<0.05$), probably due to a few objects with extremely low $f_{\rm dust}$ values in the VLA-detected RQ group. Therefore, a larger sample is needed to draw more robust conclusions.

\begin{deluxetable*}{cc|ccc|cc}[htb!]
\centering
\tablecaption{The p values of statistical tests. \label{tab:stats}}
\tablehead{
\multicolumn{2}{c|}{Group 1} & \multicolumn{3}{c|}{VLA-detected RQ} & \multicolumn{2}{c}{RL} \\
\multicolumn{2}{c|}{Group 2} & \colhead{undetected} & \colhead{RL} & \multicolumn{1}{c|}{undetected control\textsuperscript{a}} & \colhead{undetected} & \colhead{undetected control\textsuperscript{b}}
}
\startdata
SFR & KS & 0.50 & 0.97 & 0.33 & 0.93 & 0.77 \\
 & t & 0.41 & 0.54 & 0.18 & 0.88 & 0.28 \\ \hline
$M_*$ & KS & 0.94 & 0.92 & 0.81 & 0.97 & 0.98 \\
 & t & 0.71 & 0.74 & 0.99 & 0.90 & 0.93 \\ \hline
$M_{\rm dust}$ & KS & 0.000062 & 0.020 & 0.0023 & 0.49 & 0.78 \\
 & t & 0.000013 & 0.0041 & 0.00042 & 0.55 & 0.49 \\ \hline
$M_{\rm gas}$ & KS & 0.00013 & 0.039 & 0.0046 & 0.49 & 0.78 \\
 & t & 0.000021 & 0.0046 & 0.00041 & 0.60 & 0.54 \\ \hline
$M_{\rm BH}$ & KS & 0.014 & 0.95 & 0.49 & 0.17 & 0.41 \\
 & t & 0.044 & 0.67 & 0.43 & 0.39 & 0.76 \\ \hline
$L_{\rm bol}$ & KS & 0.027 & 0.091 & 0.63 & 0.19 & 0.68 \\
 & t & 0.010 & 0.017 & 0.30 & 0.33 & 0.45 \\ \hline
$M_{\rm dust}/{\rm SFR}$ & KS & 0.0052 & 0.15 & 0.28 & 0.76 & 0.60 \\
 & t & 0.052 & 0.14 & 0.65 & 0.72 & 0.72 \\ \hline 
$f_{\rm dust}$ & KS & 0.0088 & 0.30 & 0.033 & 0.97 & 0.19 \\
 & t & 0.0015 & 0.028 & 0.0054 & 0.59 & 0.60 \\ \hline
\enddata
\tablecomments{Statistical tests (KS and t-test p-values) of physical properties between the radio-undetected, VLA-detected RQ, and RL groups, including control samples. Values below 0.05 suggest significant differences ($>2\sigma$) between Group 1 and Group 2. \textsuperscript{a} The control sample selected from the radio-undetected group, which has consistent distributions of $M_{\rm BH}$ and $L_{\rm bol}$ with the VLA-detected RQ group. \textsuperscript{b} The control sample similarly selected for the RL group.}
\end{deluxetable*}

\subsection{Ionized Outflows} \label{subsec:specresult}

\begin{figure*}[htb!]
    \centering
    \includegraphics[width=0.9\linewidth]{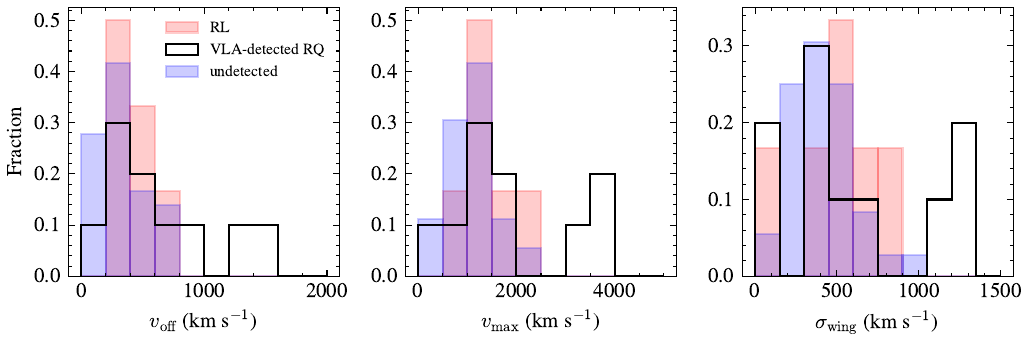}
    \caption{The distributions of $v_{\rm off}$, $v_{\max}$ and $\sigma_{\rm wing}$ of [O {\sc iii}] 5007 \text{\AA} with color coding the same as Figure \ref{fig:hist_init}.}
    \label{fig:OIII_outflow}
\end{figure*}

From the spectral fitting, we analyze the outflow properties of ionized gas in this sample. We mainly utilize the wing component of the [O {\sc iii}] emission line as a diagnostic tracer of galactic outflows. In total, 111 quasars have spectral wavelength coverage containing the [O {\sc iii}] emission line, but only 53 quasars have SNR $>2$ for both the wing component and the core component. We measure the outflow velocity $v_{\rm off}$ by the velocity offset of the wing component relative to the core component. Following \citet{OIIIvmax}, we also estimate the maximum outflow velocity by $v_{\max}=v_{\rm off}+2\sigma_{\rm wing}$, where $\sigma_{\rm wing}$ is the $\sigma$ of the wing component. Figure \ref{fig:OIII_outflow} shows the distributions of $v_{\rm off}$, $v_{\max}$ and $\sigma_{\rm wing}$ of [O {\sc iii}] 5007 \text{\AA}. From the distributions, a few quasars in the VLA-detected RQ group show outflows with very high velocities ($v_{\rm off},\ \sigma_{\rm wing}>1000\text{ km/s}$ and $v_{\max}>3000\text{ km/s}$), while the quasars in the RL group and the radio-undetected group exhibit significantly lower outflow velocities, with most systems confined to $v_{\rm off},\ \sigma_{\rm wing}<900\text{ km/s}$ and $v_{\max}<2000\text{ km/s}$.

Apart from the asymmetric [O {\sc iii}] emission line profile, some quasars in our sample show blue-shifted broad absorption line (BAL) features associated with the C {\sc iv} or Si {\sc iv} emission lines, which also indicate high-velocity outflows along the line of sight. \citet{SDSSDR16Q} identified BAL quasars (hereafter BALs) in all DR16Q quasars at $1.57\leq z \leq5.6$ by measuring the “BALnicity index” \citep[BI;][]{indexBI} and the intrinsic absorption index \citep[AI;][]{indexAI}. BI is calculated as:
\begin{equation}
    {\rm BI}=-\int_{25000}^{3000}\left[1-\frac{f(v)}{0.9}\right]C(v){\rm d}v,
\end{equation}
where $f(v)$ is the normalized flux density as a function of velocity blueshift from the C {\sc iv} emission-line center. $C(v)=1$ when $1-f(v)/0.9$ is continuously positive over a velocity interval $\Delta v\geq2000\ \rm km/s$, while $C(v)=0$ in other cases. Similarly, AI is calculated as:
\begin{equation}
    {\rm AI}=-\int_{25000}^{0}\left[1-\frac{f(v)}{0.9}\right]C(v){\rm d}v.
\end{equation}

Based on the measured BI and AI index, a discrete BAL probability {\tt BAL\_PROB} is assigned to each quasar \citep{SDSSDR16Q}. For quasars with good continuum fittings, {\tt BAL\_PROB} = 1 refers to cases where BI is more than ten times the uncertainty. All of these quasars are unambiguous BALs by visual inspection. If BI = 0, but AI is similarly significant, these quasars are assigned {\tt BAL\_PROB} = 0.9. These quasars are nearly all BALs, but their absorption trough is under 2000 km/s wide and/or extends closer to the line center than 3000 km/s. We are less confident in other quasars with {\tt BAL\_PROB} $< 0.9$, and therefore do not consider them in this work.

We examine the {\tt BAL\_PROB} value of our sample in order to obtain a more complete understanding of the outflows in addition to the [O {\sc iii}] emission line. There are a total of 208 sources with redshifts between 1.57 and 5.6, with 10 VLA-detected RQ quasars, 10 RL quasars and 182 radio-undetected quasars. The fractions of different {\tt BAL\_PROB} values in each group are listed in Table \ref{tab:BALPROB}. The fraction of {\tt BAL\_PROB} $\geq0.9$ in the VLA-detected RQ group and the RL group is similar, but much larger than that of the radio undetected group, indicating a larger fraction of the outflow hosted system in radio-detected quasars. The presence of a significant population of RL BAL quasars has also been noted in \citet{Becker2000}. Furthermore, the fraction of {\tt BAL\_PROB}=1 in the VLA-detected RQ group is higher than that of the RL group or the radio-undetected group, suggesting that the outflow-induced absorption line in the VLA-detected RQ group is wider and/or has a larger velocity offset. The fraction of BALs and the velocity of the [O {\sc iii}] emission line indicate a larger fraction of quasars in the VLA-detected RQ group hosting high-velocity outflows.

\begin{deluxetable}{cccccc}[htb!]
\centering
\tablecaption{The fraction of different {\tt BAL\_PROB} values in each group. \label{tab:BALPROB}}
\tablehead{
\colhead{Group} & \colhead{Total} & \multicolumn{4}{c}{{\tt BAL\_PROB}} \\
\colhead{} & \colhead{} & \colhead{$=1$} & \colhead{$=0.9$} & \colhead{$\geq0.9$} & \colhead{$<0.9$} 
}
\startdata
VLA-detected RQ & 10 & 40\% & 20\% & 60\% & 40\% \\
RL & 10 & 20\% & 40\% & 60\% & 40\% \\
undetected & 182 & 10.4\% & 19.8\% & 30.2\% & 69.8\% \\
\enddata
\end{deluxetable}

\section{Discussion} \label{sec:discuss}

From the SED fitting results, we find that the dust mass and the gas mass in the VLA-detected RQ group are significantly lower ($p<0.05$, Table \ref{tab:stats}) than those of the RL group and the radio-undetected group (Figure \ref{fig:hist_init}), while the three groups of quasars share similar SFRs and $M_*$. The differences remain significant after resampling to control for the consistent distributions of $M_{\rm BH}$ and $L_{\rm bol}$.

\begin{figure*}[htb!]
    \centering
    \includegraphics[width=0.8\linewidth]{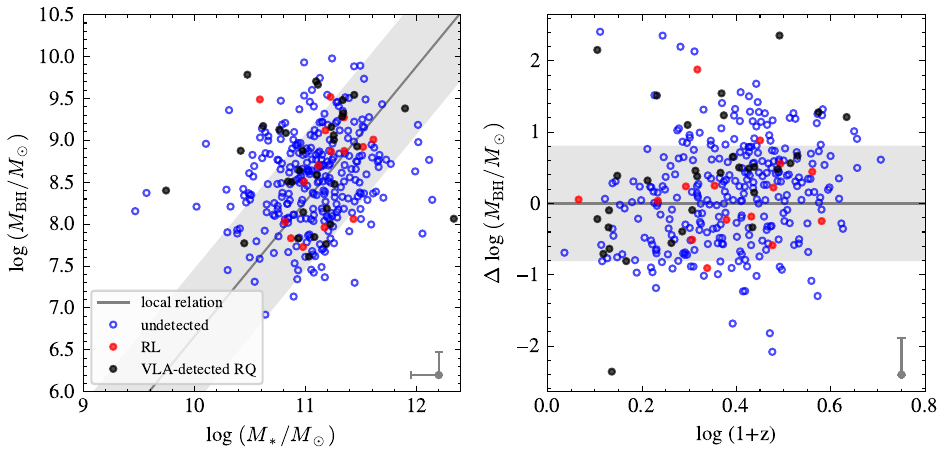}
    \caption{\textit{Left}: The distribution of our quasar sample on the $M_{\rm BH}-M_*$ plane. \textit{Right}: The offset of log $M_{\rm BH}$ with respect to the local relation \citep{Greene2020}. The RL and VLA-detected RQ groups are shown in red and black solid dots respectively. The radio-undetected group is shown in open blue circles. The gray line refers to the local relation from \citet{Greene2020} with the shaded region indicating a scatter of 0.81 dex. The typical uncertainty is presented in the lower right corner of each panel. }
    \label{fig:MBH_Mstar}
\end{figure*}

It is relevant to note that a few quasars in the VLA-detected RQ group have relatively low redshift ($z\sim0.3$, Figure \ref{fig:MBH_Mstar}, will be discussed below), which may contribute to a selection effect. Given the small sample size of this group, these low-redshift quasars constitute a significant fraction, potentially contributing to the observed lower values of $M_{\rm dust}$. To ensure consistent distributions of redshift and mitigate the potential bias from different evolutionary stages, we perform a statistical analysis on a subsample with $z>0.4$. Within this subsample, the redshift distributions across the three groups become consistent. Repeating our analysis still yields significant statistical differences in $M_{\rm dust}$ between the VLA-detected RQ group and the other two groups, with t-test p value $=2.9\times10^{-3}$ (compared with the RL group) and $1.7\times10^{-3}$ (compared with the radio-undetected group). This indicates that the observed differences in $M_{\rm dust}$ are not dominated by the presence of low-redshift quasars.

\begin{figure*}[htb!]
    \centering
    \hspace{12mm}
    \includegraphics[width=0.8\linewidth]{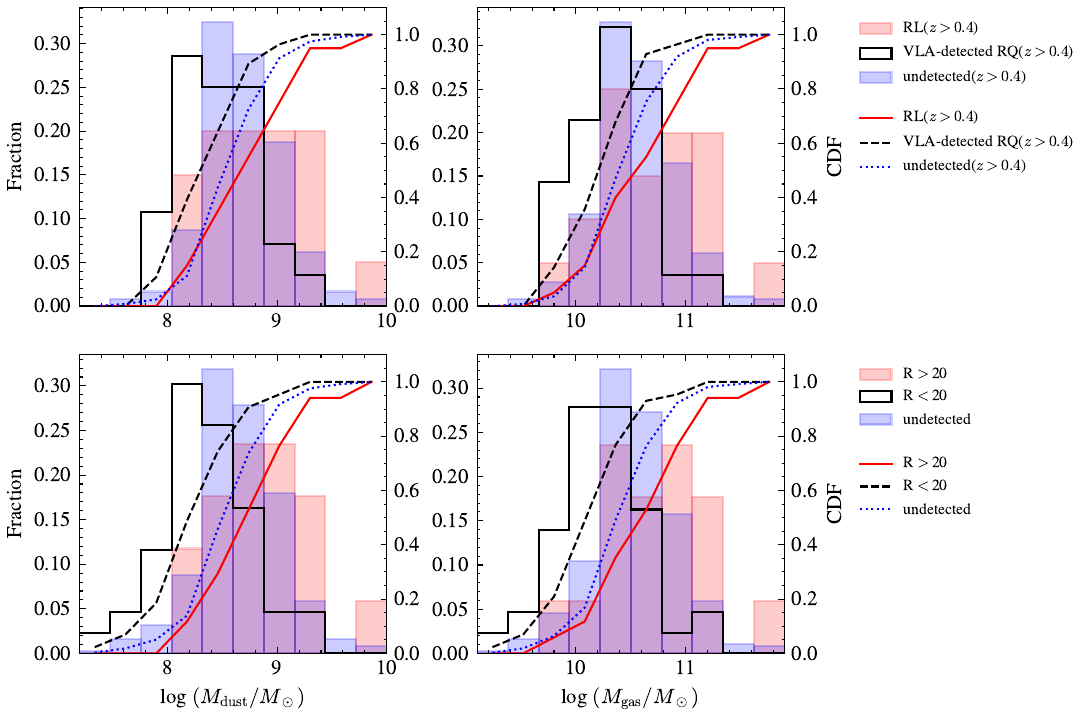}
    \caption{\textit{Top}: The distributions (histogram, left y-axis) and cumulative distribution functions (line, right y-axis) of $M\rm_{dust}$ and $M\rm_{gas}$ of the RL (red), VLA-detected RQ (black, outlined) and radio-undetected quasars (blue) with $z>0.4$. \textit{Bottom}: Same as the top panel but for the R$>$20 group (red), the R$<$20 group (black, outlined) and the radio-undetected group (blue).}
    \label{fig:hist_init_R20}
\end{figure*}

To further test the results, we adjust the radio-loudness threshold from the conventional $R = 10$ to $R = 20$, including more radio-intermediate quasars to the VLA-detected RQ group. After the reclassification, we compare the host galaxy properties among the newly defined groups (Figure \ref{fig:hist_init_R20}) and find that the differences in $M_{\rm dust}$ between VLA-detected RQ and radio-undetected quasars, as well as between VLA-detected RQ and RL quasars become slightly more significant, with lower t-test p value $=3.0\times10^{-6}$ and $1.4\times10^{-4}$ respectively. This result further supports that the host galaxies of AGNs with weak/moderate radio activities tend to have less gas content.

The origins of radio emission in RQ AGNs are of various forms, such as star formation, winds, and low-power jets \citep{Panessa2019}. As the SFR distributions of different groups are consistent, the gas deficiency observed in the VLA-detected RQ group could not be directly linked to star formation. Therefore, we mainly focus on the latter two mechanisms. 

Winds may be a possible mechanism that accounts for the moderate radio emission in RQ AGNs, as well as expels the dust and gas in their host galaxies. Multi-wavelength observations, including X-ray absorption line features and optical signatures such as broad absorption lines and [O {\sc iii}] line profiles, have revealed the ubiquitous presence of galactic winds across various scales \citep[e.g.,][]{Harrison2014}. Without exception, the quasars in our sample also exhibit outflow features such as broad absorption lines (Table \ref{tab:BALPROB}) and [O {\sc iii}] wing components (Figure \ref{fig:OIII_outflow}). 

Wind shocks can accelerate relativistic electrons and produce synchrotron radio emission, which can account for the elevated radio emission in RQ AGNs. Several studies have found positive correlations between radio emission and [O {\sc iii}] velocity width in RQ AGNs, further supporting this scenario \citep[e.g.,][]{Zakamska2014, Hwang2018}. These studies found positive relations between [O {\sc iii}] velocity width and AGN mid-infrared luminosity, indicating that the wind is fundamentally powered by the AGN radiation energy. Through quantitative analysis, a recent study by \citet{Liao2024} further strengthens this case, where they argue that the observed radio luminosities can be explained by winds with an energy conversion efficiency ($\eta=L_{\rm wind}/L_{\rm bol}$) that aligns with both previous observations and simulations.

However, if outflows powered by AGN radiation, which usually have a strong correlation with $L_{\rm bol}$ \citep[e.g.,][]{Fiore2017}, are responsible for depleting the cold gas reservoirs, one would expect to see a consequent correlation between $M_{\rm dust}$ or $M_{\rm gas}$ and $L_{\rm bol}$. Instead, we find no such correlation, i.e., the differences in $M_{\rm dust}$ and $M_{\rm gas}$ distributions across the three groups remain significant after requiring consistent distributions of both $L_{\rm bol}$ and $M_{\rm BH}$. Therefore, the wind mechanism does not appear to be enough to explain the gas deficiency in the VLA-detected RQ group.

Another possible physical mechanism to account for the moderate radio emission in RQ AGNs is the relativistic jet. Many studies with high-resolution radio observations have revealed the existence of low-power and compact radio jets in traditionally defined RQ AGNs \citep[e.g.,][]{Jarvis2019, Ulivi2024, Girdhar2024}. Furthermore, both observations \citep[e.g.,][]{Jarvis2019, Ulivi2024} and simulations \citep[e.g.,][]{Mukherjee2018, Meenakshi2022} have suggested that low-power ($<10^{44}\rm{\ erg\ s^{-1}}$) and compact jets ($<1\rm{\ kpc}$) can have a significant impact on host galaxies by injecting turbulence in the ISM and accelerating outflows. 

\begin{figure*}[htb!]
    \centering
    \includegraphics[width=\linewidth]{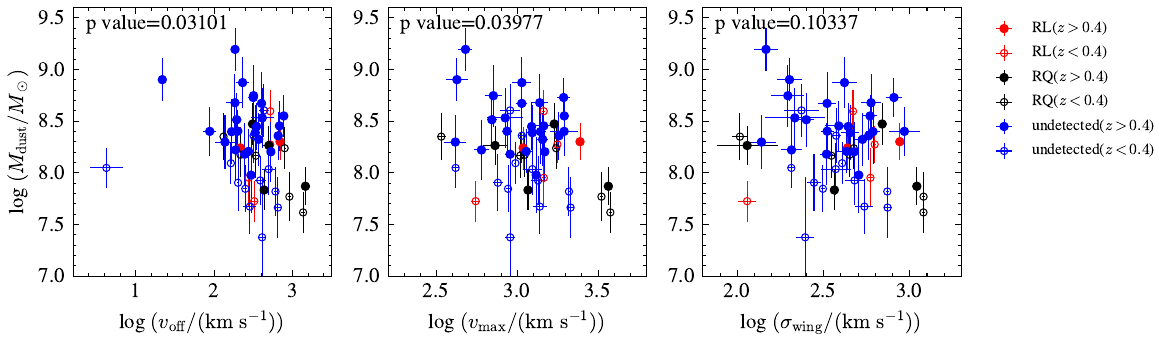}
    \caption{The relations between $M_{\rm dust}$ and $v_{\rm off}$ (left), $v_{\max}$ (middle) and $\sigma_{\rm wing}$ (right). The RL, VLA-detected RQ and radio undetected groups are shown in red, black and blue points respectively. Quasars with $z>0.4$ are shown in filled symbols and quasars with $z<0.4$ are shown in open symbols. The p values of Pearson correlation are presented at the upper left corner. The p value in the left panel is calculated after removing the quasars with $v_{\rm off}<30\rm{\ km\ s^{-1}}$.}
    \label{fig:Mdust_vout}
\end{figure*}

Recent simulations have revealed that the jet-ISM interaction depends on the properties of jets, such as the inclination angle between the jet and the galactic disk and the jet power \citep{Mukherjee2018}. Specifically, jets with high inclination angles to the galactic disk can easily escape, without significant interaction with the disk, while jets with low inclination angles can strongly and extensively perturb the ISM. \citet{Mukherjee2016} also suggested that compared with high-power jets, low-power jets are trapped in the ISM for a longer time and affect the ISM over a larger volume. From an observational perspective, \citet{Molyneux2019} explored the relation between radio size and the prevalence of extreme ionized outflows, finding the compact jets might be responsible for the stronger outflows. These results suggest that the low-power and compact jets may serve as an efficient AGN feedback mechanism that drives outflows and expels gas in host galaxies. 

The higher fraction and velocity of the outflows in the VLA-detected RQ group also support this scenario (Section \ref{subsec:specresult}). To further explore the connection between outflows and cold gas reservoirs, we investigate the relations between $M_{\rm dust}$ and the outflow properties inferred from the [O {\sc iii}] emission line. As shown in Figure \ref{fig:Mdust_vout}, we find negative correlations between $M_{\rm dust}$ and outflow velocities $v_{\rm off}$ and $v_{\max}$, though the correlations exhibit considerable scatter, likely due to diverse and complex physical conditions across our sample. In addition, the correlations still exist when only considering quasars with $z>0.4$ (filled symbols in Figure \ref{fig:Mdust_vout}). However, the correlation between $M_{\rm dust}$ and $\sigma_{\rm wing}$ is less significant (p value $=0.1$). The reason may be the fact that $\sigma_{\rm wing}$ traces the turbulence and complexity of outflows, while $v_{\rm off}$ and $v_{\rm max}$ mainly trace the bulk velocity, which should be more relevant to the expelling of dust.

However, as the [O {\sc iii}] emission line probes the ionized gas outflows, we lack data on the atomic and molecular gas outflows, which are more directly associated with the cold gas reservoirs traced by the dust emission \citep[e.g.,][]{Cicone2014,Veilleux2020}. Additionally, due to the limitations of the resolution of current radio data, most of the quasars in our sample are not resolved, restricting detailed analysis of the radio morphology, which could help differentiate between the wind and the jet.

With the measured $M_{\rm BH}$ and $M_*$, we further investigate the distributions of our FIR-bright quasar sample on the $M_{\rm BH}-M_*$ plane. From the distributions in Figure \ref{fig:MBH_Mstar}, our sample generally follows and lies slightly above the local $M_{\rm BH}-M_*$ relation \citep{Greene2020}, with a mean offset $\Delta\log M_{\rm BH}=0.17$. This offset may partly due to the redshift evolution of the $M_{\rm BH}-M_*$ relation (right panel of Figure \ref{fig:MBH_Mstar}). We find no significant differences in the distributions across the three groups across this parameter space. The lack of differences indicates that while the observed lower $M_{\rm dust}$ in the VLA-detected RQ group provides a snapshot of current cold ISM content in the host galaxies, which may lead to reduced future star formation, the cumulative feedback from past radio activity has not yet introduced significant differences to the $M_{\rm BH}-M_*$ relation. To better understand the evolution of galaxies and the interplay between feedback and mass assembly, further studies such as large-scale environments, star formation histories, as well as larger sample sizes across broader redshift ranges will be essential.

\section{Summary} \label{sec:sum}

We present a study of a FIR-bright quasar sample comprising 436 SDSS quasars spanning a wide redshift range of $0.08\leq z\leq4.1$. Utilizing multi-wavelength data from UV to radio bands, we perform SED fitting with CIGALE to derive key physical properties of the quasar host galaxies, including the SFR, $M_*$, $M_{\rm dust}$ and $M_{\rm gas}$. Additionally, by analyzing the spectral data from the SDSS database with PyQSOFit, we estimate the $M_{\rm BH}$ and the [O {\sc iii}] outflow velocity ($v_{\rm off}$ and $v_{\max}$). Our main findings are as follows.

1. Compared to RL quasars and radio-undetected quasars, VLA-detected RQ quasars exhibit significantly lower $M_{\rm dust}$ and $M_{\rm gas}$ ($\sim0.3$ dex, Figure \ref{fig:hist_init}), %$t_{\rm dep}$ and $f_{\rm dust}$ (Figure \ref{fig:tdepfdust}) 
but show consistent distributions of SFR and $M_*$. We further restrict the reduced $\chi^2$ of the SED fitting to $\chi^2_nu<5$, as suggested by \citet{reducedchi2}, and require the quasar to have a reliable BH mass measurement. Moreover, we construct two control samples, ensuring their $M_{\rm BH}$ and $L_{\rm bol}$ distributions match those of the VLA-detected RQ quasars and RL quasars, respectively. The comparisons with control samples yield consistent results, showing evidence of gas deficiency and the presence of AGN feedback in the VLA-detected RQ group.

2. A larger fraction of VLA-detected RQ quasars show evidence of ionized outflows (Table \ref{tab:BALPROB}), along with higher outflow velocities (Figure \ref{fig:OIII_outflow}). We find negative correlations between $M_{\rm dust}$ and outflow velocities including $v_{\rm off}$ and $v_{\max}$, suggesting outflows as the possible indicator of AGN feedback, that is related to the removal of cold gas reservoirs. However, we lack observational evidence of molecular outflows, which directly trace the cold gas reservoir in host galaxies. Therefore, observations of this sample using molecular gas tracers such as CO emission lines are in need.

3. The distribution of our quasar sample on the $M_{\rm BH}-M_*$ plane generally follows but lies slightly above the local relation (with a mean offset $\Delta \log M_{\rm BH}=0.17$). Additionally, we find no significant difference in the distribution across the three groups.

Our current analysis of gas, radio emission and outflows is limited to global properties, such as total gas mass and integrated radio luminosity. For a deeper insights into the specific mechanisms of AGN feedback, observations with higher resolution are essential for examining the detailed gas distribution, velocity fields and radio morphologies across different quasar groups. Additionally, observations of high-excitation molecular emission lines are necessary to investigate the gas excitation states within our sample. These detailed properties are important for understanding the impact of AGN feedback on host galaxies.

\section*{Acknowledgments}
We thank the reviewer for valuable comments that have improved the manuscript. We acknowledge support from the National Natural Science Foundation of China (NSFC) with grant Nos. 12173002, 11991052, and the Ministry of Science and Technology of the People's Republic of China with grant No. 2022YFA1602902. LCH was supported by the National Science Foundation of China (12233001) and the China Manned Space Program (CMS-CSST-2025-A09). We acknowledge support from ANID programs BASAL CATA FB210003 (ET, FEB), FONDECYT Regular 1241005  (FEB, ET) and 1250821 (ET, FEB), and Millennium Science Initiative AIM23-0001 (FEB).

Funding for the Sloan Digital Sky Survey IV has been provided by the Alfred P. Sloan Foundation, the U.S. Department of Energy Office of Science, and the Participating Institutions. SDSS-IV acknowledges support and resources from the Center for High Performance Computing  at the University of Utah. The SDSS website is www.sdss4.org.

SDSS-IV is managed by the Astrophysical Research Consortium for the Participating Institutions of the SDSS Collaboration including the Brazilian Participation Group, the Carnegie Institution for Science, Carnegie Mellon University, Center for Astrophysics | Harvard \& Smithsonian, the Chilean Participation Group, the French Participation Group, Instituto de Astrof\'isica de Canarias, The Johns Hopkins University, Kavli Institute for the Physics and Mathematics of the Universe (IPMU) / University of Tokyo, the Korean Participation Group, Lawrence Berkeley National Laboratory, Leibniz Institut f\"ur Astrophysik Potsdam (AIP),  Max-Planck-Institut f\"ur Astronomie (MPIA Heidelberg), Max-Planck-Institut f\"ur Astrophysik (MPA Garching), Max-Planck-Institut f\"ur Extraterrestrische Physik (MPE), National Astronomical Observatories of China, New Mexico State University, New York University, University of Notre Dame, Observat\'ario Nacional / MCTI, The Ohio State University, Pennsylvania State University, Shanghai Astronomical Observatory, United Kingdom Participation Group, Universidad Nacional Aut\'onoma de M\'exico, University of Arizona, University of Colorado Boulder, University of Oxford, University of Portsmouth, University of Utah, University of Virginia, University of Washington, University of Wisconsin, Vanderbilt University, and Yale University.

This paper makes use of the following ALMA data: ADS/JAO.ALMA\#2022.1.00029.S. ALMA is a partnership of ESO (representing its member states), NSF (USA) and NINS (Japan), together with NRC (Canada), NSTC and ASIAA (Taiwan), and KASI (Republic of Korea), in cooperation with the Republic of Chile. The Joint ALMA Observatory is operated by ESO, AUI/NRAO and NAOJ. 
Some/all of the data presented in this paper were obtained from the Mikulski Archive for Space Telescopes (MAST) at the Space Telescope Science Institute. The specific observations analyzed can be accessed via \dataset[https://doi.org/10.17909/T9H59D]{https://doi.org/10.17909/T9H59D}. STScI is operated by the Association of Universities for Research in Astronomy, Inc., under NASA contract NAS5–26555. Support to MAST for these data is provided by the NASA Office of Space Science via grant NAG5–7584 and by other grants and contracts. This research has made use of the NASA/IPAC Extragalactic Database, which is funded by the National Aeronautics and Space Administration and operated by the California Institute of Technology.

The Herschel spacecraft was designed, built, tested, and launched under a contract to ESA managed by the Herschel/Planck Project team by an industrial consortium under the overall responsibility of the prime contractor Thales Alenia Space (Cannes), and including Astrium (Friedrichshafen) responsible for the payload module and for system testing at spacecraft level, Thales Alenia Space (Turin) responsible for the service module, and Astrium (Toulouse) responsible for the telescope, with in excess of a hundred subcontractors.
AllWISE makes use of data from WISE, which is a joint project of the University of California, Los Angeles, and the Jet Propulsion Laboratory/California Institute of Technology, and NEOWISE, which is a project of the Jet Propulsion Laboratory/California Institute of Technology. WISE and NEOWISE are funded by the National Aeronautics and Space Administration.
The National Radio Astronomy Observatory and Green Bank Observatory are facilities of the U.S. National Science Foundation operated under cooperative agreement by Associated Universities, Inc. All the AllWISE data used in this paper can be found via \dataset[https://doi.org/10.17909/10.26131/IRSA1]{https://doi.org/10.26131/IRSA1}.

\vspace{5mm}
\software{Astropy \citep{Astropy2013},  
          CASA \citep{CASA},
          Matplotlib \citep{Matplotlib2007}
          Numpy \citep{Numpy2011},
          Scipy \citep{Scipy2020}.
          }
% \facilaties{}

\appendix

\section{Dust Mass Measurements} \label{app:dust}

In addition to performing UV-to-radio SED fitting using CIGALE with multiple degree-of-freedom templates, we also employed the MBB model to fit the sub-mm photometry to estimate the dust mass, which offers greater simplicity with fewer free parameters. The MBB model is a widely used approach to characterize galactic dust emission in the sub-mm bands. This model assumes thermal emission from dust grains to be a single-temperature blackbody modified by a frequency-dependent emissivity. For the optically thin case, the form of the MBB model can be expressed as:
\begin{equation}
    S_\nu=\frac{1+z}{D_L^2}\kappa_0\left(\frac{\nu}{\nu_0}\right)^\beta B_\nu(T_{\rm dust})M_{\rm dust},
\end{equation}
where $D_L$ is the luminosity distance, $\kappa_0$ is the emissivity per unit dust mass at frequency $\nu_0$, $\beta$ is the dust emissivity index, $B_\nu(T_{\rm dust})$ is the Planck function at dust temperature $T_{\rm dust}$, and $M_{\rm dust}$ is the dust mass. For a comparison with CIGALE-derived dust mass, we adopt $\kappa_{0,1200\mu \rm m}=0.4 \rm\ cm^2\ g^{-1}$ and $\nu_0=c/1200\mu\rm m$. For radio-detected quasars, we include a power-law component $S_\nu=f_0(\nu/\nu_0)^{-\alpha}$ to account for the radio emission that may contribute to the observed sub-mm flux. For simplicity, the AGN torus contribution is directly subtracted using the best-fit CIGALE AGN model to mitigate the contamination from AGNs at short wavelength.

\renewcommand\thefigure{A\arabic{figure}}
\setcounter{figure}{0}
\begin{figure}[htb!]
    \centering
    \includegraphics[width=0.4\linewidth]{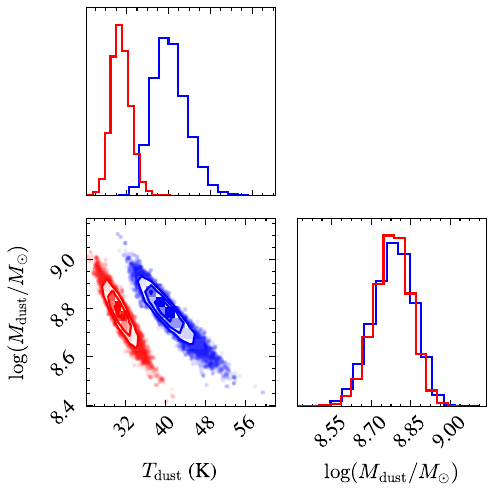}
    \includegraphics[width=0.4\linewidth]{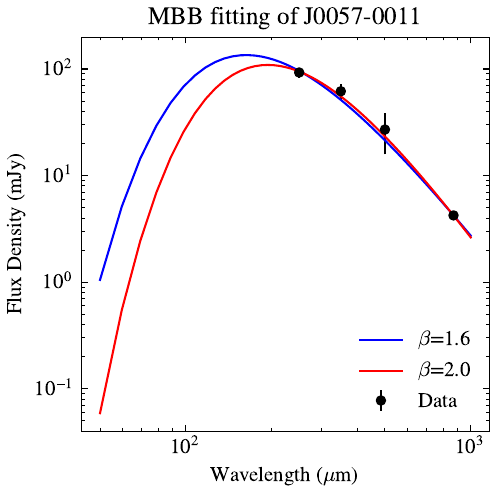}
    \caption{The MBB fitting results of SDSS J005743.78-001158.9. The result of $\beta=1.6$ is shown in blue and the result of $\beta=2$ is shown in red.}
    \label{fig:example_MBB}
\end{figure}

\begin{figure}[htb!]
    \centering
    \includegraphics[width=0.6\linewidth]{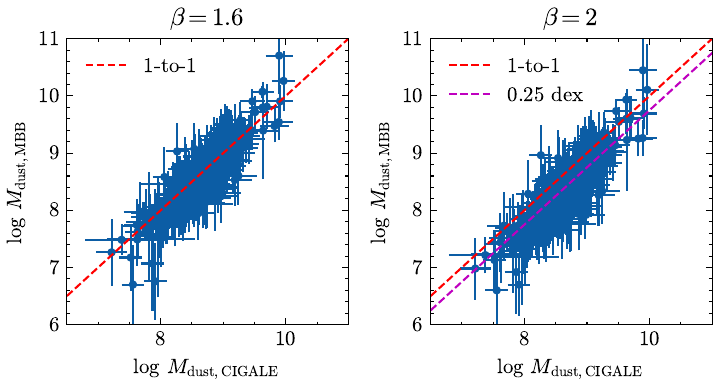}
    \caption{The comparisons of dust mass estimated from CIGALE and MBB fitting. The left panel shows the MBB fitting results with $\beta$ fix to 2 and the right panel shows the results with $\beta=1.6$. The red dashed line is the one-to-one ratio line and the magenta dashed line is offset by 0.25 dex.}
    \label{fig:MBB}
\end{figure}

The model typically requires three free parameters: $T_{\rm dust}$, $\beta$ and $M_{\rm dust}$, and two additional parameters $f_0$ and $\alpha$ for radio-detected objects. Given that $\beta$ is usually poorly constrained with limited sub-mm photometric data points, we fix $\beta$ to 1.6 and 2.0 due to its degeneracy with $T_{\rm dust}$, then examine how $\beta$ affects the dust mass measurements. To avoid overfitting, we fix $T_{\rm dust}=40 \rm\ K$ when only two data points are available in sub-mm bands and fix $\alpha=0.7$ with single radio data point. To quantify the uncertainties of parameters, we employ the Markov Chain Monte Carlo (MCMC) method to obtain probability distributions. The fitting example and results are shown in Figure \ref{fig:example_MBB} and Figure \ref{fig:MBB}.

From the comparisons, the dust masses measured from CIGALE $M_{\rm dust, CIGALE}$ are consistent with those measured from MBB fitting with $\beta=1.6$ [$M_{\rm dust, MBB}(\beta=1.6)$], while $M_{\rm dust, CIGALE}$ are $\sim$0.25 dex larger than $M_{\rm dust, MBB}(\beta=2)$, indicating that the dust mass measurement depends on our choice of parameters. In addition to the dust emissivity index $\beta$, the choice of different $\kappa_0$ values will also affect the derived dust mass. However, such systematic offsets do not affect our main conclusion that relative differences in dust mass remain significant across groups with varying radio loudness. For a better constrain on the dust and gas masses, observations such as more FIR-to-millimeter photometric data points and molecular gas emission lines are needed.

\bibliography{HerS_VLA}{}
\bibliographystyle{aasjournal}

\end{document}